\documentclass{elsart}

\usepackage[square,comma]{natbib}
\usepackage{graphicx}
\usepackage{pxfonts}
\usepackage{lineno}

\usepackage{amssymb}

\date{}
\journal{}

\begin{document}

\thispagestyle{empty}
\begin{Large}
\textbf{DEUTSCHES ELEKTRONEN-SYNCHROTRON}

\textbf{\large{Ein Forschungszentrum der
Helmholtz-Gemeinschaft}\\}
\end{Large}

DESY 10-165

September 2010

\begin{eqnarray}
\nonumber &&\cr \nonumber && \cr \nonumber &&\cr
\end{eqnarray}
\begin{eqnarray}
\nonumber
\end{eqnarray}
\begin{center}
\begin{Large}
\textbf{Way to increase the user access at the LCLS baseline}
\end{Large}
\begin{eqnarray}
\nonumber &&\cr \nonumber && \cr
\end{eqnarray}

\begin{large}
Gianluca Geloni,
\end{large}
\textsl{\\European XFEL GmbH, Hamburg}
\begin{large}

Vitali Kocharyan and Evgeni Saldin
\end{large}
\textsl{\\Deutsches Elektronen-Synchrotron DESY, Hamburg}
\begin{eqnarray}
\nonumber
\end{eqnarray}
\begin{eqnarray}
\nonumber
\end{eqnarray}
ISSN 0418-9833
\begin{eqnarray}
\nonumber
\end{eqnarray}
\begin{large}
\textbf{NOTKESTRASSE 85 - 22607 HAMBURG}
\end{large}
\end{center}
\clearpage
\newpage

\begin{frontmatter}



\title{Way to increase the user access at the LCLS baseline}


\author[XFEL]{Gianluca Geloni\thanksref{corr},}
\thanks[corr]{Corresponding Author. E-mail address: gianluca.geloni@xfel.eu}
\author[DESY]{Vitali Kocharyan}
\author[DESY]{and Evgeni Saldin}

\address[XFEL]{European XFEL GmbH, Hamburg, Germany}
\address[DESY]{Deutsches Elektronen-Synchrotron (DESY), Hamburg,
Germany}

\begin{abstract}
Although the LCLS photon beam is meant for a single user, the baseline undulator is long enough to serve two users simultaneously. To this end, we propose a setup composed of two simple elements:
an X-ray mirrors pair for X-ray beam deflection, and a short ($4$m-long) magnetic chicane, which creates an offset for
mirrors pair installation in the middle of the baseline undulator. The insertable mirrors pair can be used for spatial separation
of the X-ray beams generated in the first and in the second half of the baseline undulator. The method of deactivating one half and
activating another half of the undulator is based on the rapid switching of the FEL amplification process. As proposed elsewhere, using a kicker installed upstream of the LCLS baseline undulator and an already existing corrector in the first half of the undulator, it is possible to rapidly switch the X-ray beam from one user to another, thus providing two active beamlines at any time. We present simulation results dealing with the LCLS baseline, and show that it is possible to generate two saturated SASE X-ray beams in the whole 0.8 - 8 keV photon energy range in the same baseline undulator. These can be exploited to serve two users. Implementation of the proposed technique does not perturb the baseline mode of operation of the LCLS undulator.  Moreover, the magnetic chicane setup is very flexible, and can be used as a self-seeding setup too. We present simulation results for the LCLS baseline undulator with SHAB (second harmonic afterburner) and show that one can produce monochromatic radiation at the 2nd harmonic as well as at the 1st.  We describe an efficient way for obtaining multi-user operation at the LCLS hard X-ray FEL. To this end, a photon beam distribution system based on the use of crystals in the Bragg reflection geometry is proposed. The reflectivity of crystal deflectors can be switched fast enough by flipping the crystals with piezoelectric devices similar to those for X-ray phase retarders at synchrotron radiation facilities.  It is then possible to distribute monochromatic hard X-rays among $6$ independent experiments, thereby enabling $6$ users to work in parallel in the near and far experimental halls.
\end{abstract}

\end{frontmatter}



\section{\label{sec:intro} Introduction}

The success of LCLS \cite{LCLS2} motivated the planning of further development. According to this planning, the two new undulators
will be installed in a new undulator tunnel, and their radiation be directed to novel underground experimental halls. The LCLS baseline undulator will remain operative at least during the next decade \cite{WUTAL}.

As a result, enhancing the operating LCLS baseline capability becomes a challenging problem, subject to many constraints including low cost, little available time to perform changes (and to be subtracted from operations) and almost no risk i.e. guarantee of safe return to the baseline mode of operation. In a previous paper of us \cite{OURL} we began to analyze possibilities for enhancing the LCLS baseline subject to the above constraints. The setup proposed in that paper is based on the insertion of a single, short magnetic chicane setup, which allows one to implement a novel single-bunch self-seeding scheme exploiting a single crystal monochromator, which is extremely compact and can be straightforwardly installed in the LCLS baseline undulator. The setup can be installed in little time, and is not expensive. With the radiation beam monochromatized down to the Fourier transform limit, a variety of very different techniques leading to further improvements of the LCLS performance become feasible. In particular, in \cite{OURL} we described an efficient way for obtaining full polarization control at the LCLS hard X-ray FEL. We also proposed to exploit crystals in the Bragg reflection geometry as movable deflectors for the LCLS X-ray transport systems. The hard X-ray beam can be deflected of an angle of order of a radian without perturbations. The monochromatization of the output radiation constitutes the key for reaching such result.

\begin{figure}[tb]
\includegraphics[width=1.0\textwidth]{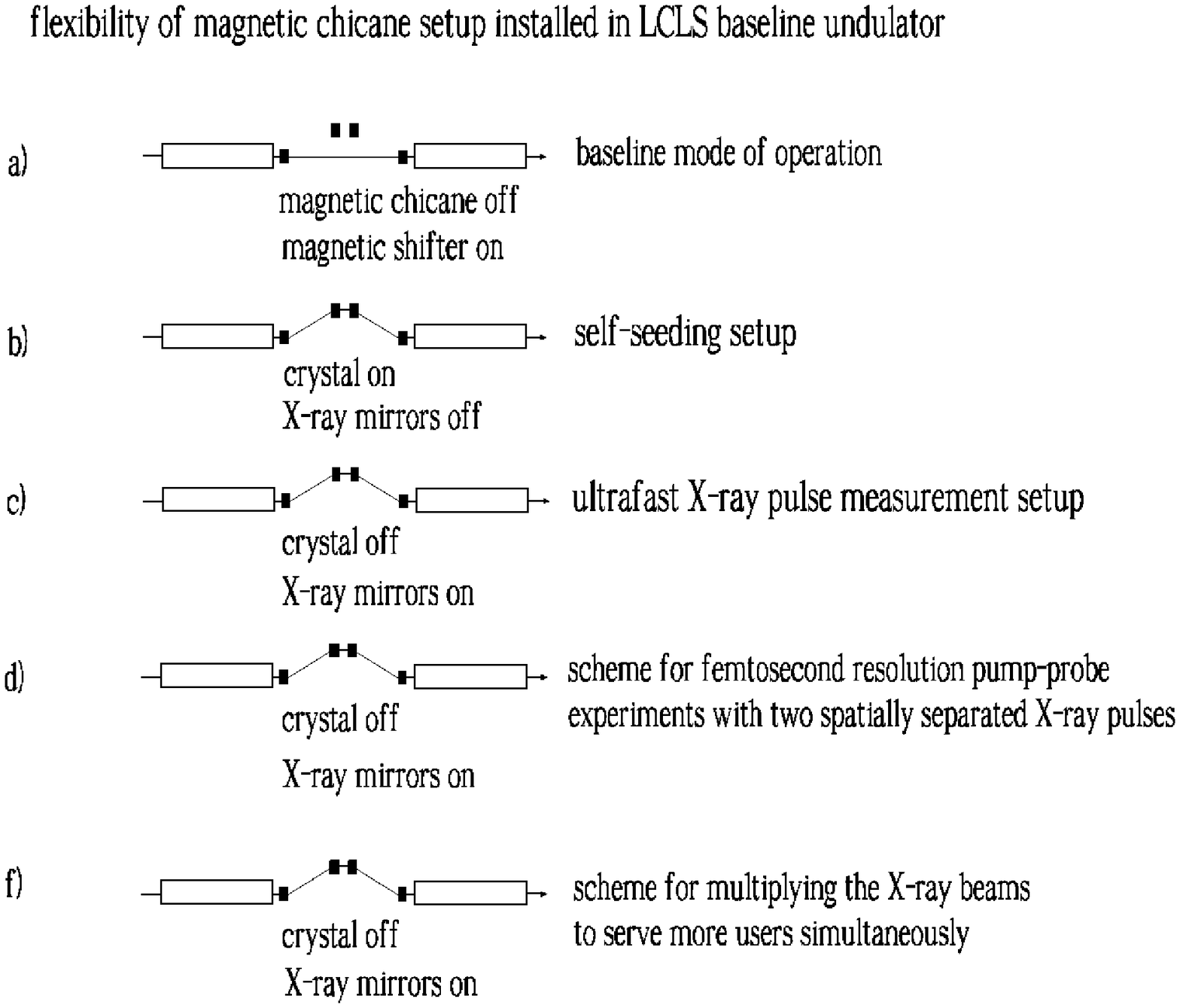}
\caption{Possible usage of the short (4 m - long) magnetic chicane setup installed in the LCLS baseline undulator} \label{lclsbm1}
\end{figure}

With reference to Fig. \ref{lclsbm1} one can appreciate the flexibility of our setup. If the chicane is switched off, the baseline mode of operation is immediately restored, Fig. \ref{lclsbm1} (a). The self-seeding scheme can be implemented by introducing a wake monochromator \cite{OURL,OURX,OURY2,OURY3}, Fig. \ref{lclsbm1} (b). By retracting the crystal, inserting X-ray mirrors in its place, and keeping the chicane switched on, one may still implement a method for ultrafast photon pulse measurements \cite{OURL, OUR05}, Fig. \ref{lclsbm1} (c). The same setup can be used to implement \cite{OURL} a pump-probe scheme which is not affected by the jitter between pump laser and X-ray pulse, Fig. \ref{lclsbm1} (d).

In this paper we will discuss how to further exploit the installation of the chicane. In particular, the setup with mirrors and chicane can be used to provide a method to serve more users simultaneously, Fig. \ref{lclsbm1} (e). Moreover, we will discuss how to further exploit the self-seeding scheme, in combination with a Second Harmonic After Burner (SHAB).

\begin{figure}[tb]
\includegraphics[width=1.0\textwidth]{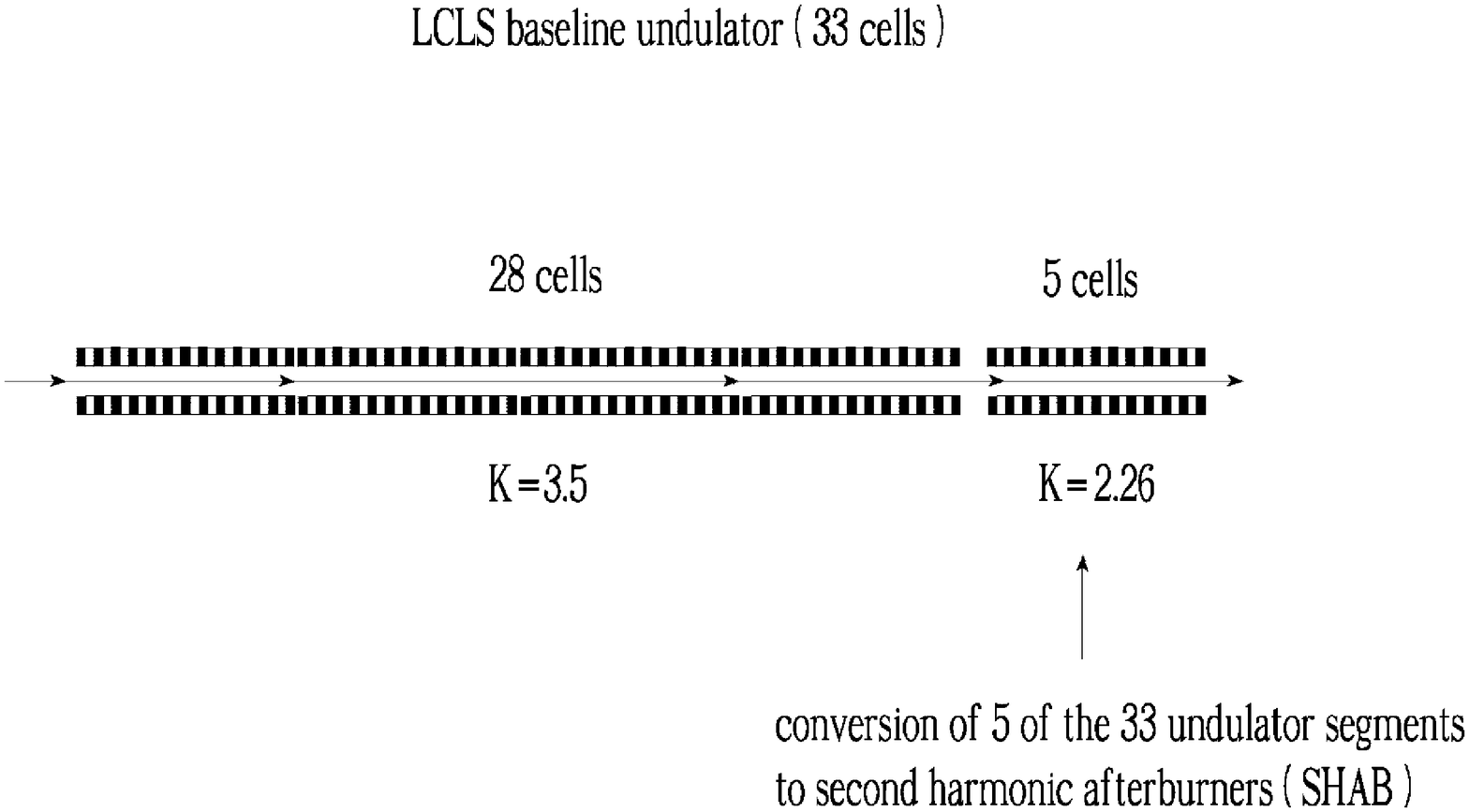}
\caption{Current design of the LCLS baseline undulator system} \label{lclsbm4}
\end{figure}
\begin{figure}[tb]
\includegraphics[width=1.0\textwidth]{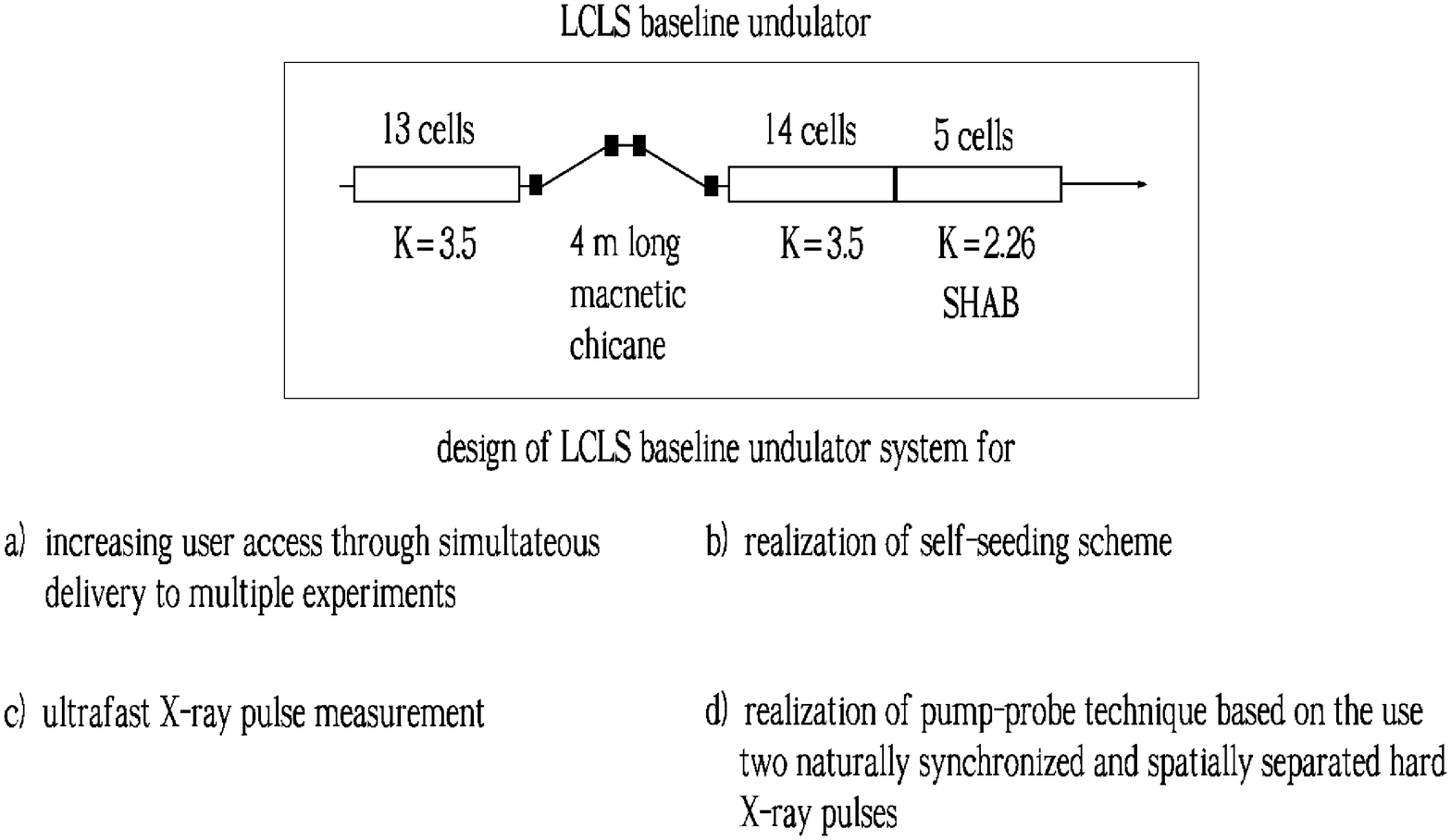}
\caption{Proposed design of the LCLS baseline undulator system} \label{lclsbm2}
\end{figure}
The LCLS team plans \cite{WUTAL} to convert the last five undulator segments of the LCLS baseline undulator to SHAB, Fig. \ref{lclsbm4}. We propose to remove\footnote{The extra undulator segment may be in principle reinstalled at the beginning of the undulator, although this may turn out difficult due to the particular geometry of the LCLS baseline. In our simulations we used a conservative approach and assumed that one undulator segment is simply removed.} the 14th undulator segment, and to substitute it with a weak chicane, Fig. \ref{lclsbm2}.

One extra use of the setup in Fig. \ref{lclsbm2}, aside for those already described in Fig. \ref{lclsbm1} (a)-(d), consist in increasing the user access at the LCLS baseline, which constitutes a very important goal. In its essence, the method is very simple. We propose to generate two SASE X-ray beams in different parts of the baseline undulator, and spatially separate them with the help of X-ray mirrors introduced in the offset provided by the chicane. This will allow to simultaneously supply two experiments. The LCLS baseline undulator is long enough to provide two SASE X-ray beams at saturation, and very small X-ray mirrors can be used to spatially separate the two X-ray beams.

The gap of the LCLS baseline undulator can only be adjusted within a few percent, and therefore the X-ray beam doubler can only operate with coupled wavelengths. Normally, if the SASE FEL operates at saturation, the quality of the electron bunch is too bad for the generation of SASE radiation in a subsequent part of undulator,
which is resonant at nearly the same wavelength. We propose to use a method for activating or deactivating SASE undulators, based on a rapid switching of the FEL amplification process in order to obtain a user doubler \cite{TESL, KICK,OUR04}.  In the LCLS baseline case
a betatron SASE switcher \cite{KICK}  may be practically advantageous. The method is discussed in the following Section \ref{increase}.

\begin{figure}[tb]
\includegraphics[width=1.0\textwidth]{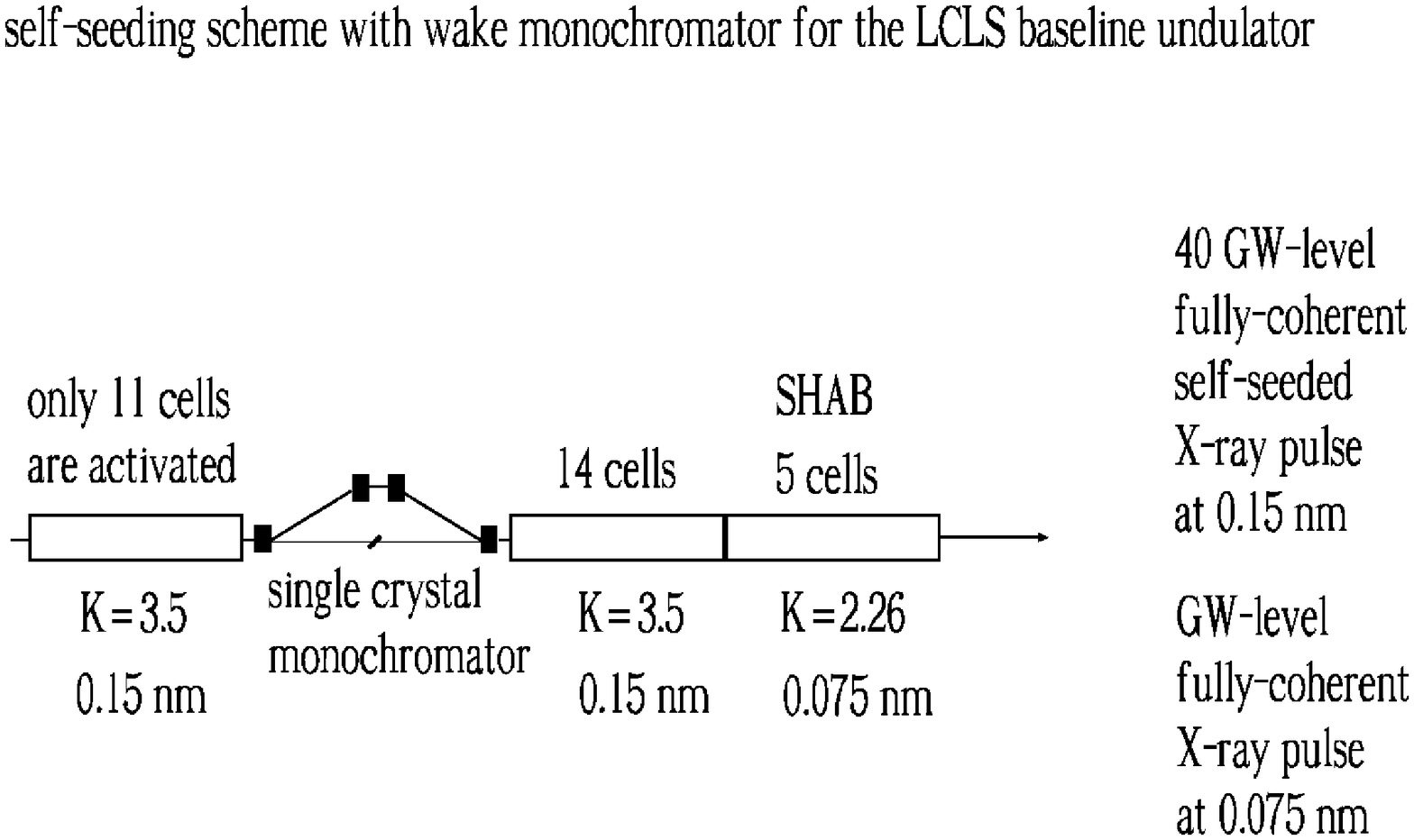}
\caption{Design of the LCLS baseline undulator system for generating highly monochromatic hard X-ray beams at the first and at the second harmonic.} \label{lclsbm3}
\end{figure}
The presence of the SHAB can be further exploited together with the self-seeding technique according to the scheme in Fig. \ref{lclsbm3}. In this setup, the first two segment of the first undulator part, before the chicane are deactivated to allow lasing in the linear regime. We performed simulations and studied the properties of the 2nd harmonic of the output radiation, which is also monochromatic due to the implementation of the self-seeding technique. The setup output will therefore consist of two monochromatic X-ray beams, at the first and at the second harmonic. Monochromaticity is the key for implementing multi-user operation in the hard X-ray range, which can be granted by using crystal deflectors and small absorption of the radiation
at $16$ keV in crystals. We suggest to flip crystals for switching of reflectivity very much like polarization-switching techniques with X-ray phase retarders at synchrotron radiation facilities, based on the use of piezoelectric components. The method is discussed in Section \ref{increase}.

After Section \ref{increase}, in Section \ref{feas} and in Section \ref{SHAB}, we study feasibility with the help of simulations based on the code Genesis \cite{GENE}. Finally, in Section \ref{conc}, we come to conclusions.

\section{\label{increase} Scheme for multiplying the X-ray beams to serve more users simultaneously}

As discussed in the Introduction, substituting a single undulator segment in the LCLS baseline with a weak chicane yields a dramatic increase of the LCLS capabilities, allowing for many different schemes to be realized without risks for the baseline mode of operation. In this Section we discuss the principles of a method to increase user access in the full photon energy range, based on SASE radiation, and to obtain a multi-user facility in hard X-ray range, based on self-seeded radiation.

\begin{figure}[tb]
\includegraphics[width=1.0\textwidth]{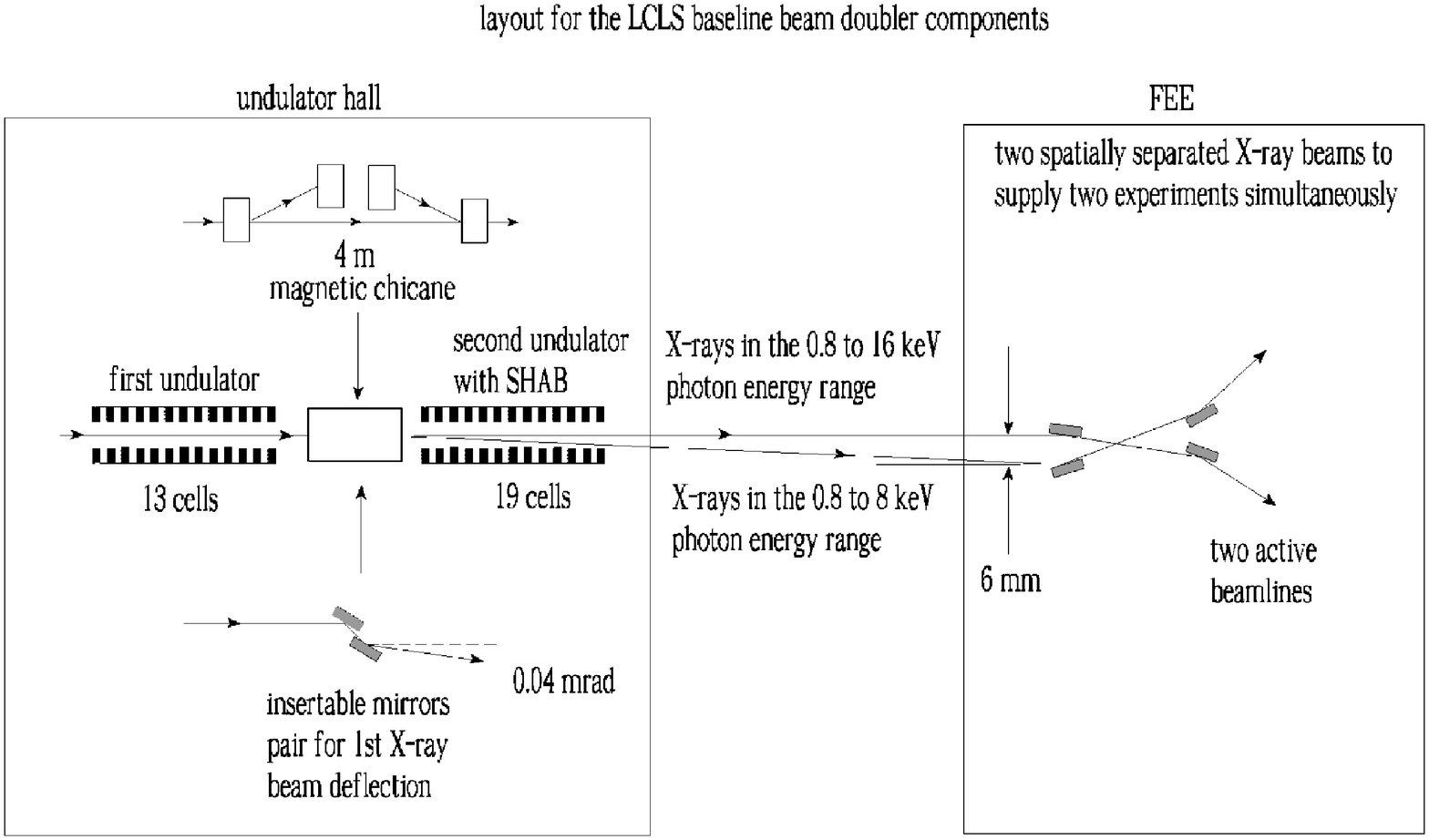}
\caption{Possible extension of the number of user stations which can operate simultaneously at the LCLS baseline.} \label{lclsbm6}
\end{figure}

\begin{figure}[tb]
\includegraphics[width=1.0\textwidth]{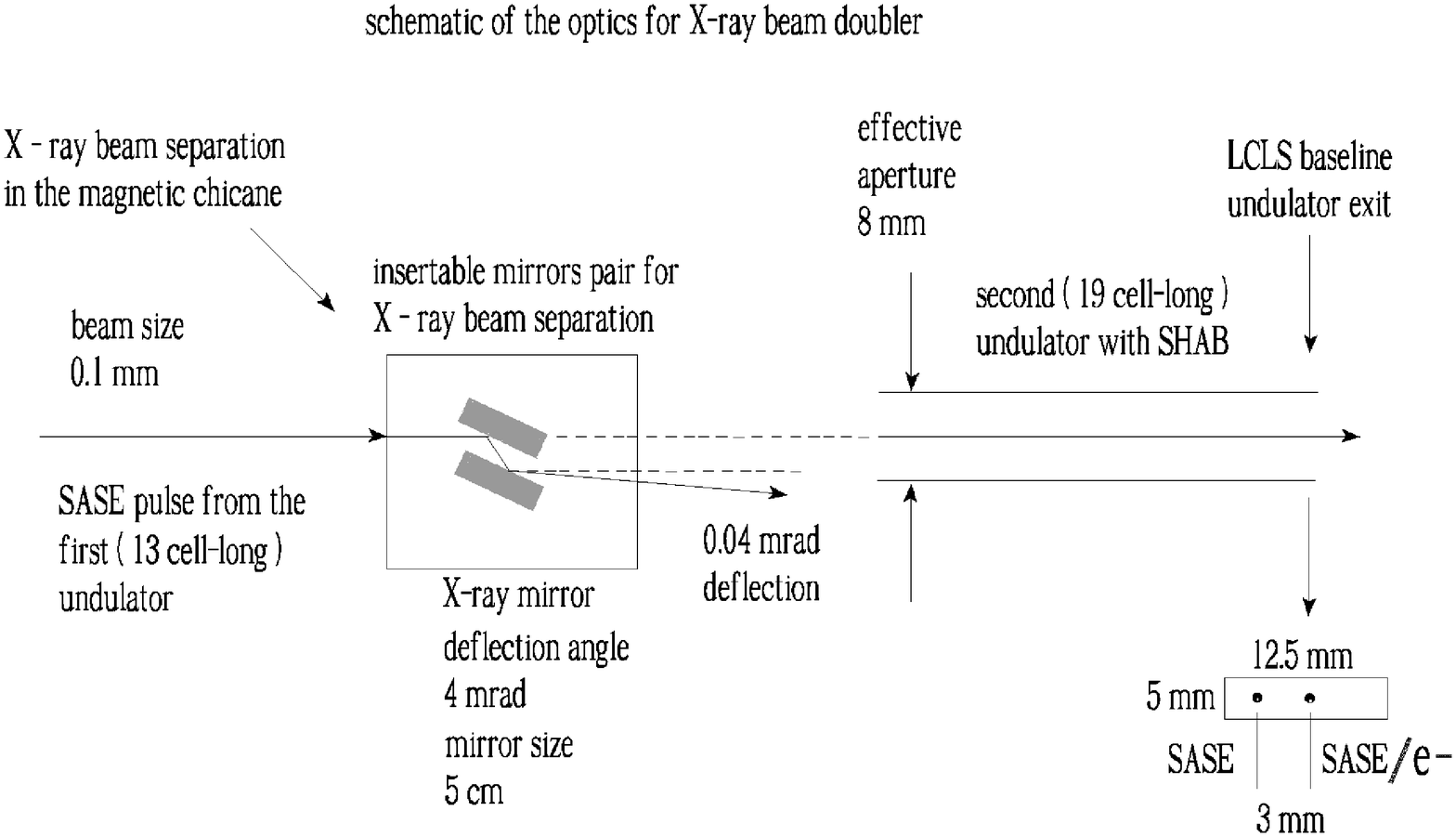}
\caption{Scheme for horizontally separating the first SASE pulse with respect to the second within the LCLS baseline undulator.} \label{lclsbm5}
\end{figure}

\begin{figure}[tb]
\includegraphics[width=1.0\textwidth]{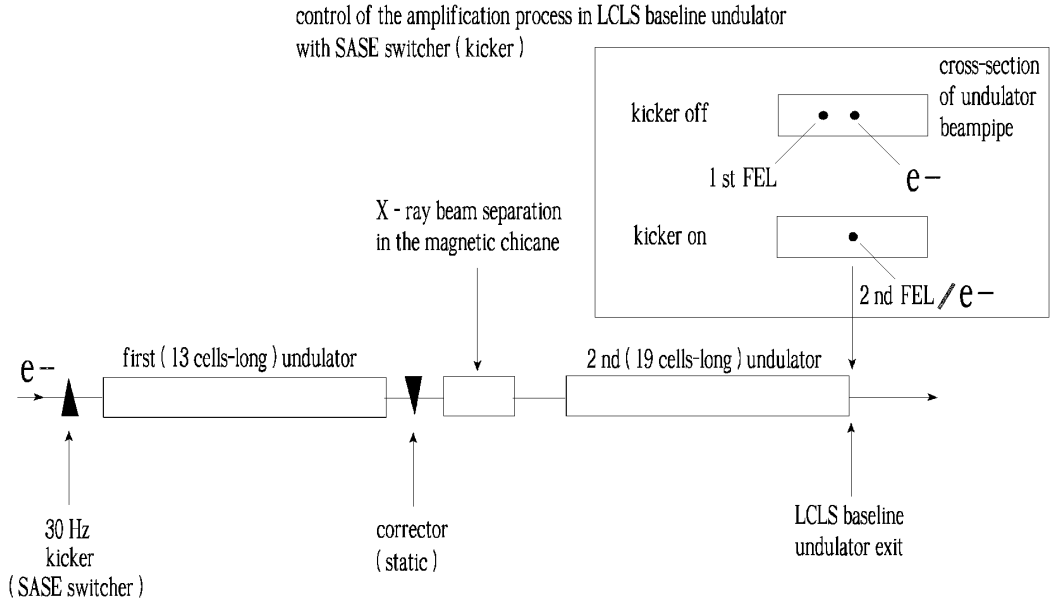}
\caption{The two modes of operation for the SASE switcher in the LCLS baseline undulator.} \label{lclsbm7}
\end{figure}
The method to increase the user access in the full photon energy range, based on SASE radiation, is sketched in Fig. \ref{lclsbm6}-\ref{lclsbm7}, and is justified by the length of LCLS baseline undulator, which is about two times longer than what is needed to reach saturation at $0.15$ nm wavelength \cite{LCLS2}. The basic idea is to exploit this extra-available length and to obtain two user simultaneous operation at LCLS baseline. The two SASE X-ray beams, generated from two different parts of the baseline undulator are spatially separated with the help of X-ray mirrors. The idea is sketched in Fig. \ref{lclsbm6}.  A sketch of the two X-ray mirrors
installed within the magnetic chicane after the first (13 cells-long) part of undulator is shown in Fig. \ref{lclsbm5}. This insertable  mirrors pair can be used to generate a deflection angle of $0.04$ mrad, which does not lead to interception of the FEL beam with the vacuum chamber within the undulator, but creates a separation of about $6$ mm in the Front End Enclosure (FEE) tunnel\footnote{Note that using a single mirror would not work. In fact, assuming a transverse size of the X-ray beam of about $0.1$mm, the footprint on a single mirror at glancing angle of $0.02$ mrad would be about $5$ m, too long to be accomodated within the chicane. The two mirrors in Fig. \ref{lclsbm5} are at a glancing angle of $2$ mrad. This yields a footprint of just $5$ cm. Then, in order to introduce a deflection of $0.04$ mrad, one only needs to rotate one mirror with respect to the other of such small amount.}. FEE mirror systems will subsequently direct the two spatially separated X-ray beams into two experimental stations.

Change of the the active part of the undulator is based on rapid switching of the FEL amplification process. For the LCLS baseline
undulator, we propose to use a method to control the SASE amplification process  based on betatron switcher \cite{KICK}\footnote{The idea of using a betatron switcher has been independently proposed at SLAC, \cite{WUTAL}.}.  Using a kicker installed upstream of the LCLS baseline undulator and existing correctors located at the end of the first part of the undulator, it is possible to rapidly switch the X-ray beam from one user to another, thus providing two active beamlines at any time, Fig. \ref{lclsbm7}. The location of the kicker must be chosen such that the betatron phase advance between kicker and corrector is  $\pi$, or a multiple thereof.  The fast kicker operates at a repetition rate of $30$ Hz, yielding a kick of about 0.01 mrad to every second bunch. Such kick is compensated by using a corrector (which always gives the same, static kick). After the corrector, any bunch which was initially  kicked goes into a straight section, and lases up to saturation in the second part of undulator. In the first part of undulator instead, the trajectory of such any bunch strongly deviates from the straight path, thus inhibiting the lasing process. When the fast kicker is off, the bunch lases in the first part of the undulator up to saturation. As a result, two SASE beams can be generated from different parts of the undulator with repetition rate $30$ Hz for each independent user. When the kicker is off, in the second undulator we have a perturbation of the orbit due to the corrector, but this is not important because in this case we already have saturation in the first undulator. In reference \cite{KICK}, a setup with three or more undulators was considered.  In our case, where we discuss only about two undulator parts, we deal with much simpler situation. We also need to have a betatron phase shift of $\pi$ between kicker and corrector, but in our scheme this is not a constraint for the undulator lengths, because the switcher scheme can always be implemented with the proper choice of the kicker position upstream of the baseline undulator.

An idea for obtaining a user doubler at LCLS II was
presented in \cite{WUTAL}.  There it is propose to tilt the second part of the undulator for obtaining spatial separation between the X-ray beams generated in the first and in the second part of the new soft X-ray undulator. Our idea to use an insertable mirrors pair within a short magnetic chicane for the same purpose, has the advantages that the first and the second part of the undulator can be kept straight, and that can therefore be applied to the LCLS baseline. Obviously, such method can be applied to the LCLS II design as well.

If the X-ray mirrors are substituted by a wake monochromator, the self-seeding scheme described in \cite{OURL} is applied, Fig \ref{lclsbm1}. In this case, there exists an efficient way for obtaining a multi-user facility at LCLS baseline in hard X-ray range, based on self-seeded radiation. To this end, we propose a photon beam distribution system based on the use of crystals in the Bragg reflection geometry as deflectors. It can be possible to deflect the full radiation pulse of an angle of the order of a radian without perturbations. The proposed photon beam distribution system would allow to switch the hard X-ray beam quickly between $6$ experimental stations in the near and far experimental halls in order to make an efficient use of the LCLS source. The high monochromatization of the output radiation constitutes the key for reaching such results. In \cite{OURL} we described self-seeding setup for LCLS baseline. In its simplest configuration, a self-seeded FEL consists of an input and an output undulator separated by a monochromator.  A distinguishing feature of our method is that it uses a single crystal in Bragg transmission geometry as monochromator, installed within the magnetic chicane.

The method has the potential for a decrease of the bandwidth of the output radiation down to the Fourier transform limit, given by the finite duration of the pulse. As discussed in the introduction, the LCLS baseline undulator design with magnetic chicane  proposed in Fig. \ref{lclsbm1} is very flexible and can be used for monochromatization purpose  too, Fig. \ref{lclsbm3}. Successful operation of the self-seeding XFEL requires the fulfillment of
several requirements. The first undulator part must operate in the deep linear regime, and not at saturation. Calculations show that in order not to spoil the electron beam quality, the number of cells in the first part of undulator should be decreased from $13$ down to $11$. This can be done simply by detuning $2$ cells, using the $2 \%$-adjustable gap of the LCLS baseline undulator. In section \ref{SHAB} we will demonstrate that for the undulator equipped with SHAB it is possible decrease the bandwidth of the second harmonoc radiation down to the transform limit as well.

\begin{figure}[tb]
\includegraphics[width=1.0\textwidth]{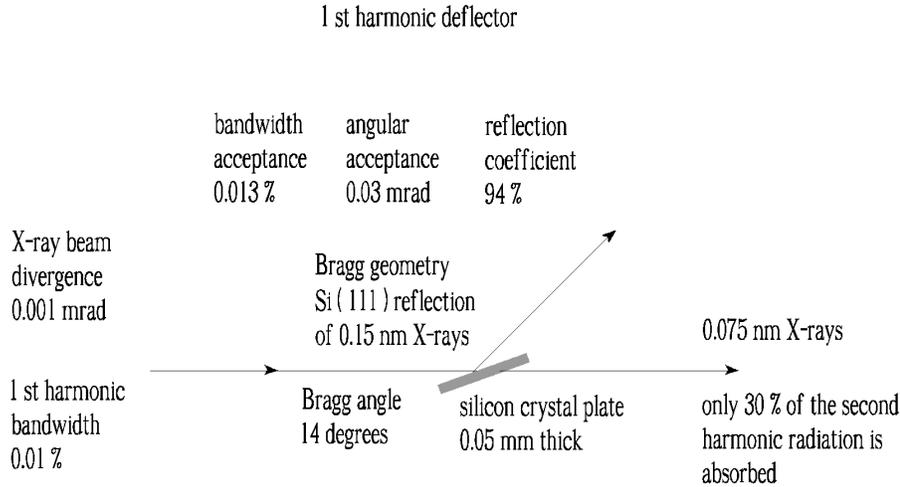}
\caption{Concept of the first harmonic photon beam deflector based on the use of a crystal in Bragg reflection geometry.} \label{lclsbm9}
\end{figure}

\begin{figure}[tb]
\includegraphics[width=1.0\textwidth]{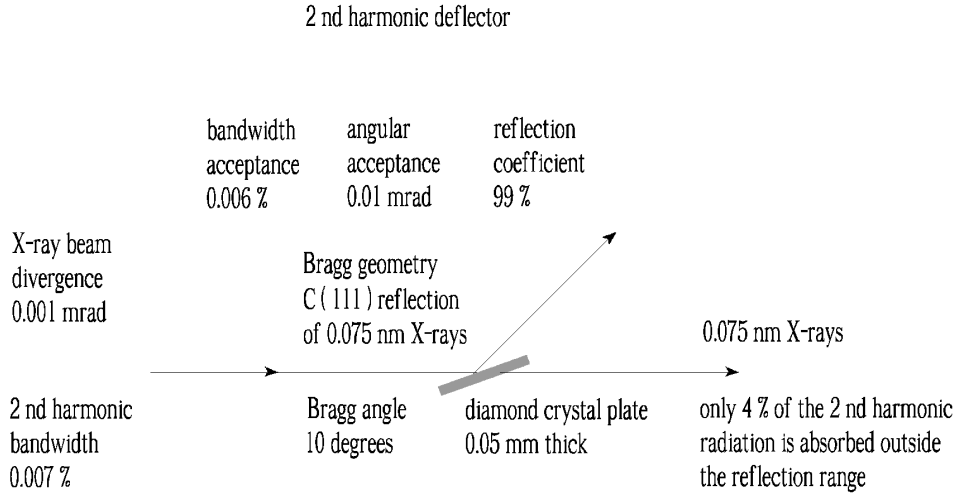}
\caption{Concept of the second harmonic photon beam deflector based on the use of a crystal in Bragg reflection geometry. High energy (2 nd harmonic) X-rays are associated with a large penetration power, which makes it possible to use relatively thick diamond deflectors. The thickness of one crystal is $0.05$ mm. The deflector can be simply switched off by tilting, which results in a change in the angle of incidence of the radiation. The angular change necessary for switching on and off the reflected beam is less than $0.1$ mrad. Only $4\%$ of the  incoming 2 nd harmonic radiation  is absorbed in this case.} \label{lclsbm10}
\end{figure}
Dealing with monochromatic radiation one may use crystals as deflector, for both the first and the second harmonic, Fig. \ref{lclsbm9} and Fig. \ref{lclsbm10}. The first harmonic deflector can be constituted of a silicon crystal, exploiting the 111
reflection in the symmetric Bragg geometry. The thickness of the first harmonic deflector is 0.05 mm. The second harmonic deflector is diamond plate of 0.05 mm thickness. It is used in Bragg reflection geometry with the C(111) reflection plane. One of the most advantages of crystal deflector in Bragg geometry is that the reflected beam is switched only by changing angle. The typical angular change necessary for the switching is less than $0.1$ mrad. This opens up a new possibility of fast switching of the reflectivity, which is necessary for multi-user operation. In order to achieve a stable photon beam deflection the rotation error must be less than $0.01$ mrad. Existing technology enables rotating crystals to satisfy these requirements. For example, at synchrotron radiation facilities, X-ray phase retarder crystals are driven by a piezoelectric device operated at hundred Hz repetition rate, which flips the crystals with a rotation error of about a fraction of a microradian \cite{HIRA2}.

\begin{figure}[tb]
\includegraphics[width=1.0\textwidth]{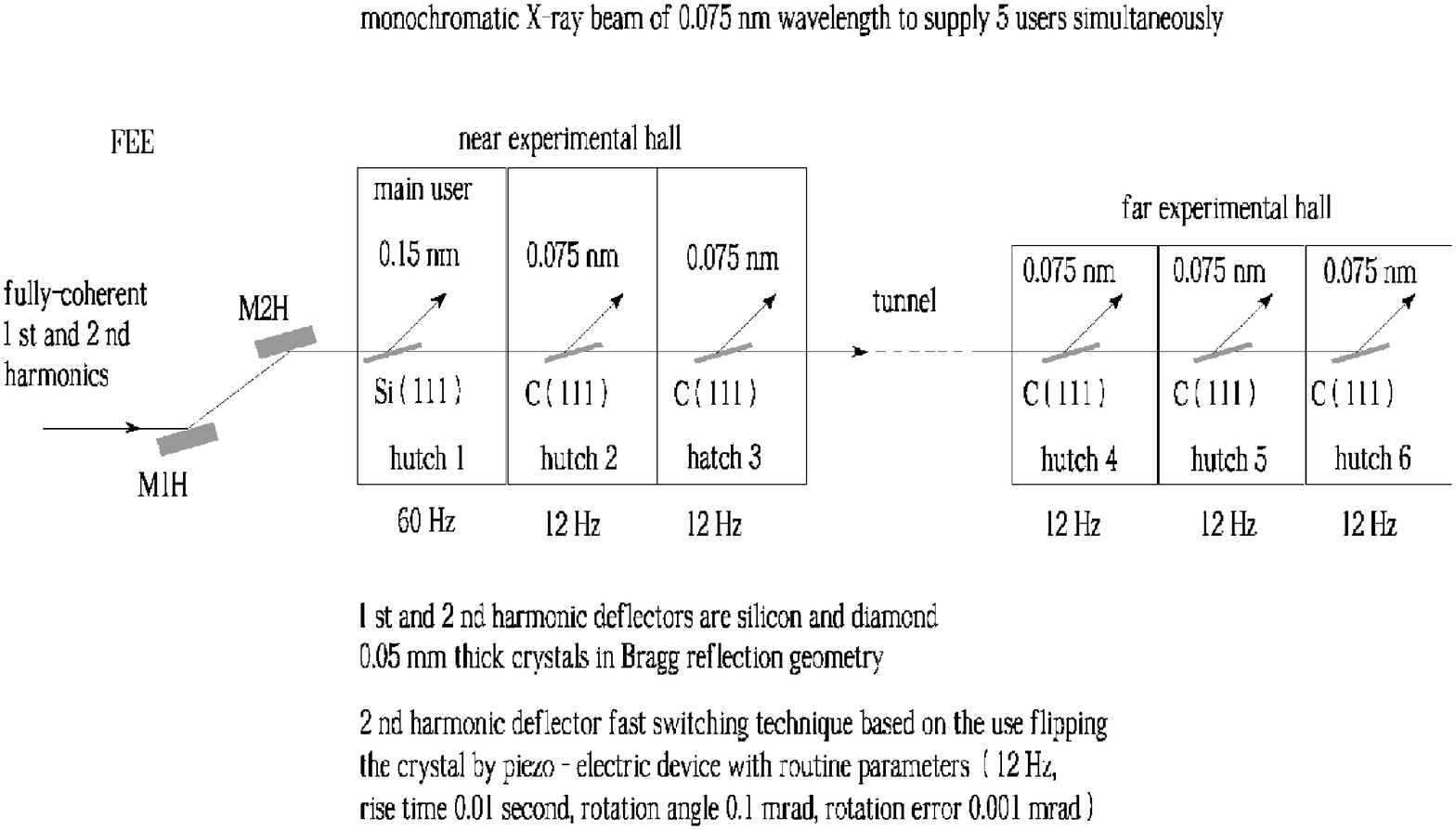}
\caption{Proposed hard X-ray LCLS baseline undulator beam line. A photon beam distribution system based on flipping crystals can provide an efficient way to obtain a multi-user facility. The monochromatization of the output first and second harmonic radiation constitutes the key for reaching such result.} \label{lclsbm8}
\end{figure}
We propose to consider a multi-user distribution system based on
(flipping) crystal deflectors, which will allow  all existing user stations to work in parallel with the same  high-quality
X-ray beams, Fig. \ref{lclsbm8}. The initial second harmonic photon beam at 0.075 nm wavelength  will thus be transformed into 5 different beams. The switching crystals need to flip at a frequency of $12$ Hz, so that each user (except the first, which receives the full $60$ Hz X-ray beam at the first harmonic) receives 12 X-ray beams per second.

\section{\label{feas} Feasibility study for SASE beam doubler}

We first consider the scheme in Fig. \ref{lclsbm6}. A pair of X-ray mirrors is inserted in the transverse offset generated by the magnetic chicane, as described in Section \ref{increase}. This feasibility study is performed with the help of the FEL code GENESIS 1.3 \cite{GENE}, running on a parallel machine, for the low-charge mode of operation ($0.02$ nC) corresponding to a bunch length of about $6$ fs. Parameters are presented in Table \ref{tt1}.

\begin{table}
\caption{Parameters for the low-charge mode of operation at LCLS used in
this paper.}

\begin{small}\begin{tabular}{ l c c}
\hline & ~ Units &  ~ \\ \hline
Undulator period      & mm                  & 30     \\
K parameter (rms)     & -                   & 2.466  \\
Wavelength            & nm                  & 0.15   \\
Energy                & GeV                 & 13.6   \\
Charge                & nC                  & 0.02 \\
Bunch length (rms)    & $\mu$m              & 1    \\
Normalized emittance  & mm~mrad             & 0.4    \\
Energy spread         & MeV                 & 1.5   \\
\hline
\end{tabular}\end{small}
\label{tt1}
\end{table}
We study, separately, the output from the first undulator, from the second, and finally, from the SHAB, constituted by the last five undulator segments.

\subsection{Output from the first undulator}

\begin{figure}[tb]
\includegraphics[width=1.0\textwidth]{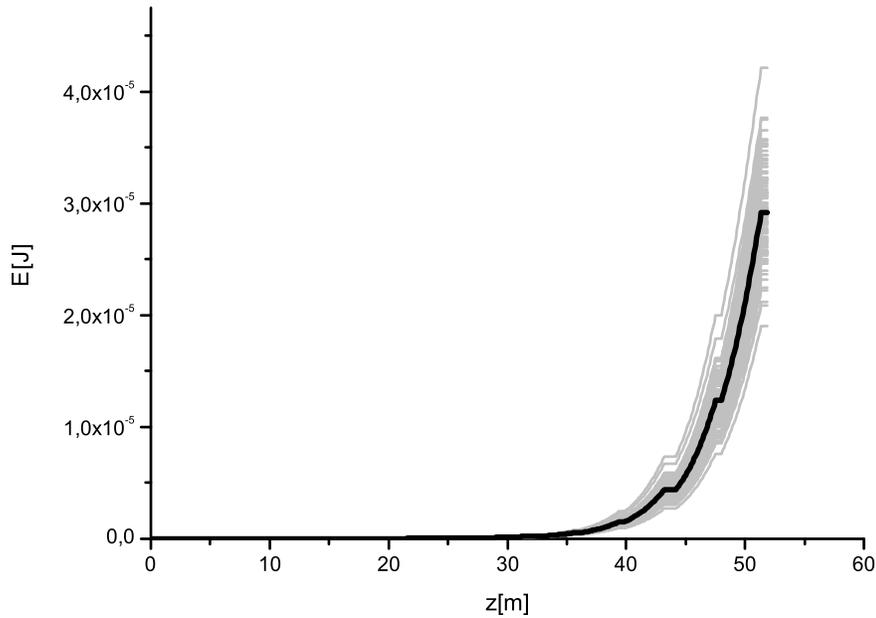}
\caption{Beam doubler setup. Energy in the X-ray radiation pulse versus the length of the first output undulator. Here the output undulator
is 13 cells-long. Grey lines refer to single shot realizations, the black line refers to the average over a hundred realizations.} \label{us1ene}
\end{figure}

\begin{figure}[tb]
\includegraphics[width=1.0\textwidth]{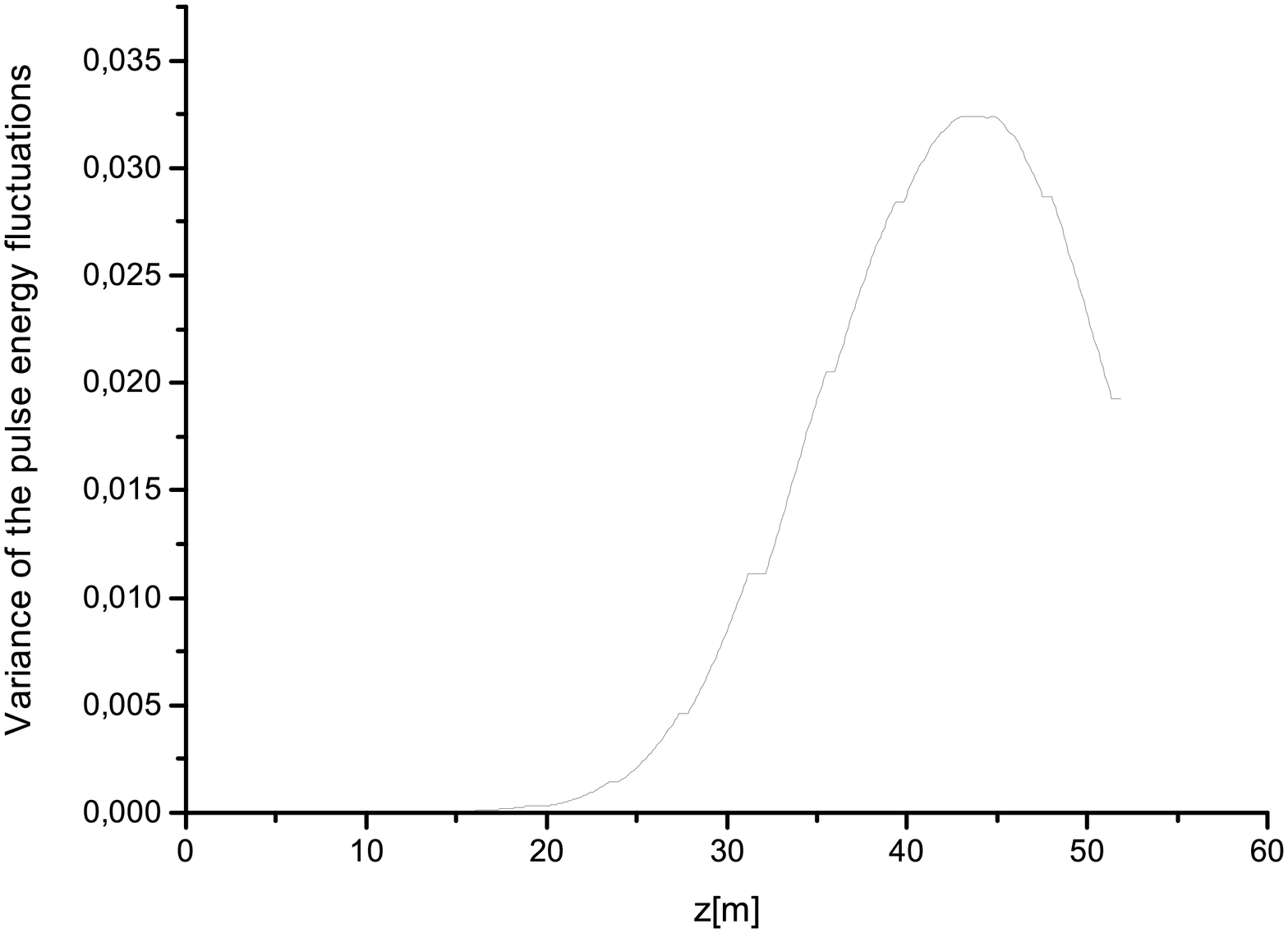}
\caption{Beam doubler setup. Variance of the energy deviation from the average as a function of the
distance inside the first output undulator. Here the output undulator is 13 cells-long.} \label{us1rms}
\end{figure}

\begin{figure}[tb]
\includegraphics[width=1.0\textwidth]{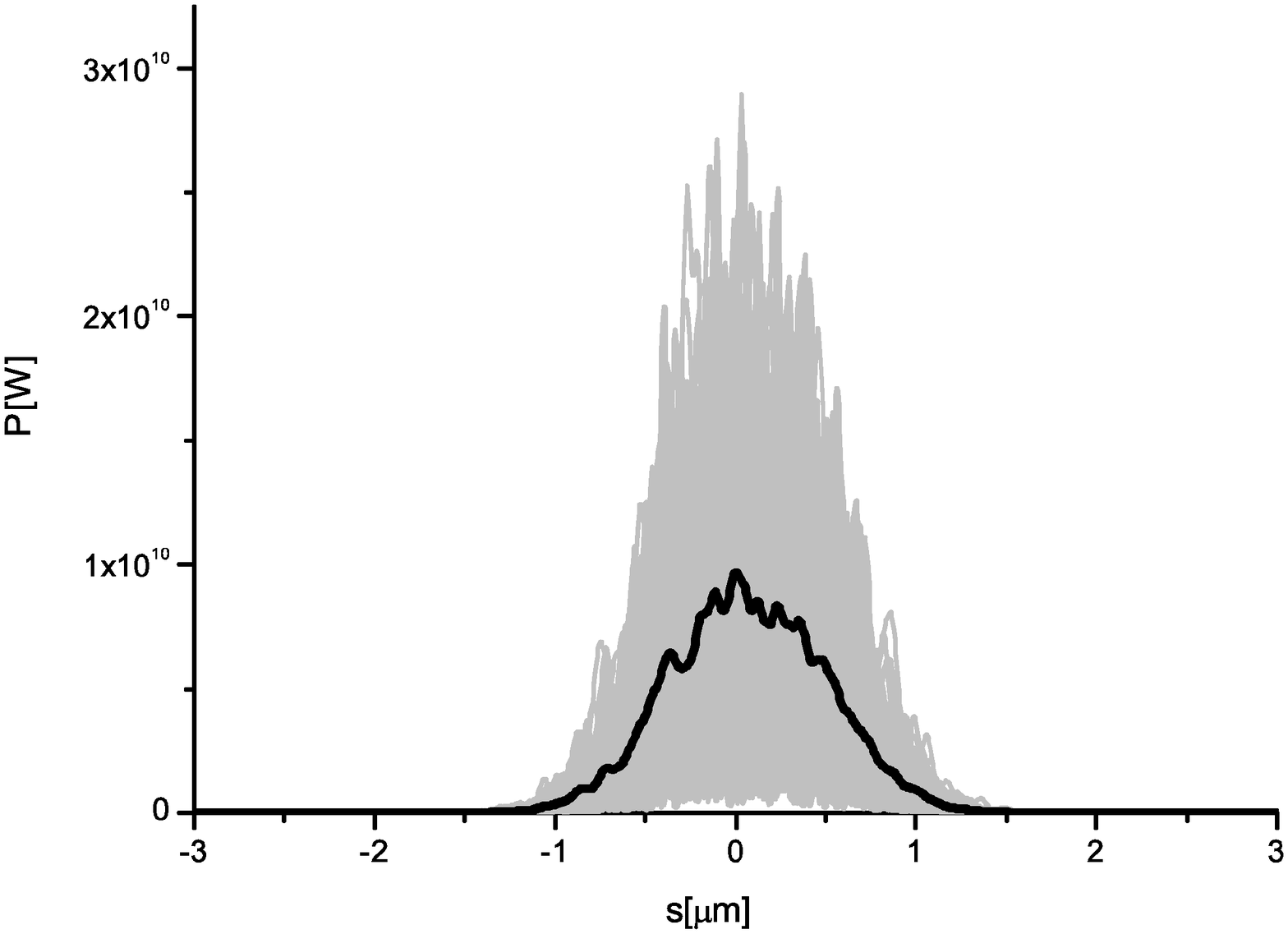}
\caption{Beam doubler setup. Power distribution after the first SASE undulator (13 cells). Grey lines refer to single shot realizations, the black line refers to the average over a hundred realizations. } \label{us1pow}
\end{figure}

\begin{figure}[tb]
\includegraphics[width=1.0\textwidth]{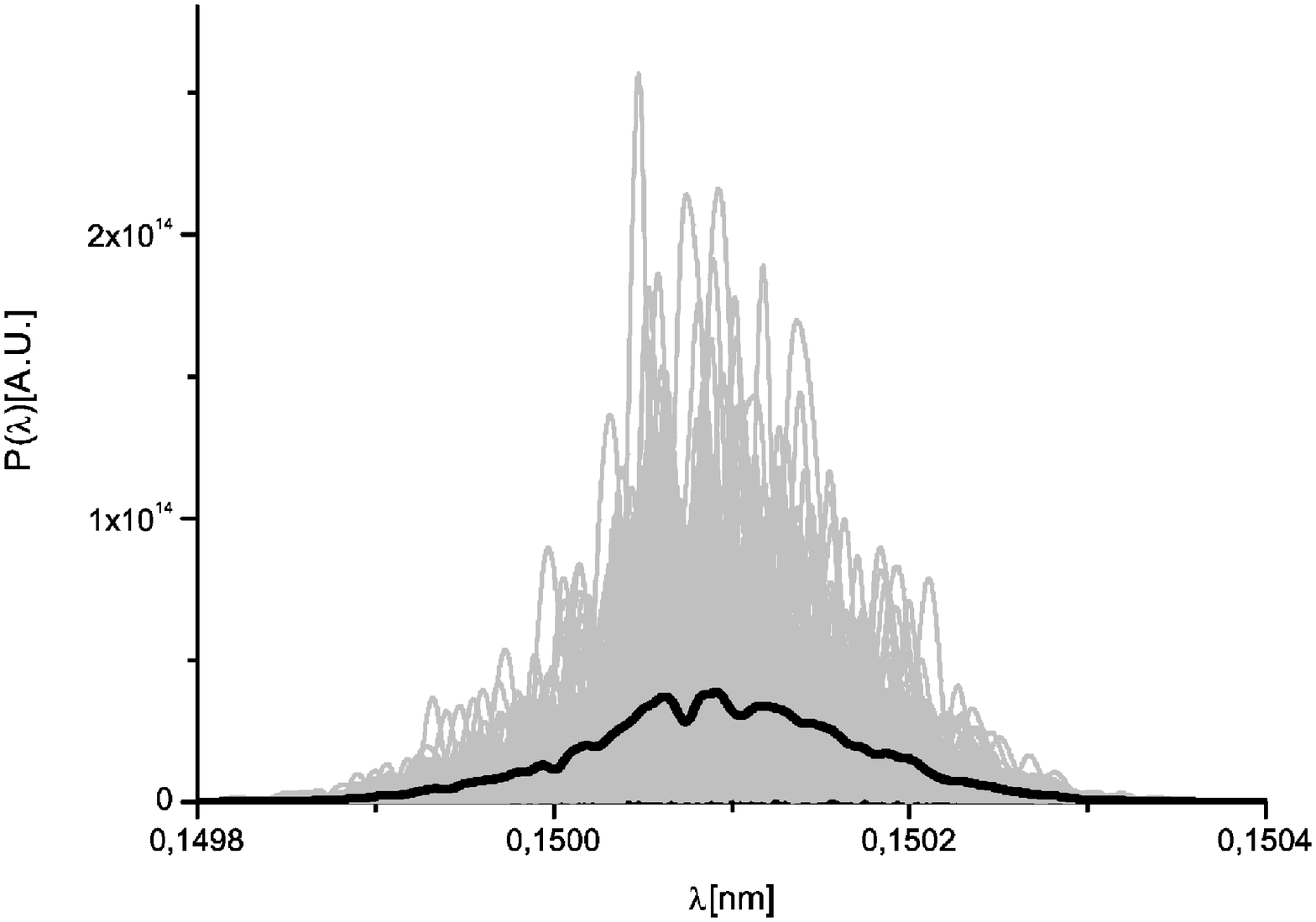}
\caption{Beam doubler setup. Spectrum of the X-ray pulse after the first SASE undulator (13 cells). Grey lines refer to single shot realizations, the black line refers to the average over a hundred realizations.} \label{us1spec}
\end{figure}

We first let the electron beam through the first undulator. Results are summarized in Fig. \ref{us1rms}, Fig. \ref{us1ene}, Fig. \ref{us1pow} and Fig. \ref{us1spec}.

Fig. \ref{us1spec} shows a typical SASE spectrum at 0.15 nm. Powers are in the $10$ GW-level, Fig. \ref{us1pow}, with energies per pulse in the order of a few tens of microjoules. The poor longitudinal coherence of the SASE pulse results in small energy fluctuations, as can be seen from Fig. \ref{us1rms}.  Fig. \ref{us1rms} also shows that the end of the linear regime is reached since, at it can be seen by inspection, the energy fluctuations drop.



\subsection{Output from the second undulator}

In the proposed scheme for the X-ray beam doubler we assume a fresh bunch at the entrance of the second undulator. We only account for an increase of the energy spread in the electron bunch due to
quantum fluctuations during the pass through the first part of
the undulator. The output characteristics of the first harmonic for the second user are calculated at 0.15 nm for a 14 cells-long undulator. The results, as can be expected, are similar as for
the first user.

\begin{figure}[tb]
\includegraphics[width=1.0\textwidth]{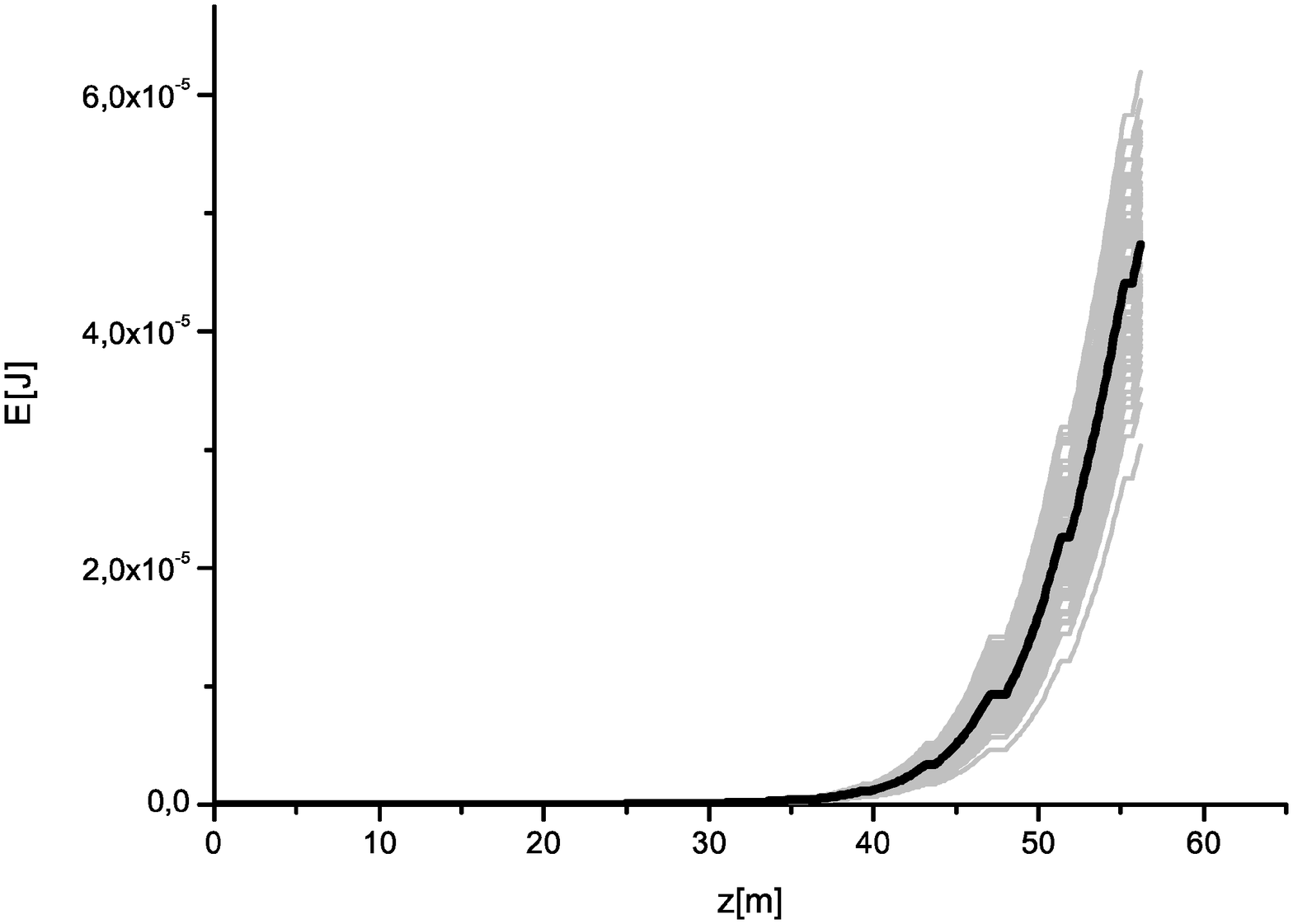}
\caption{Beam doubler setup. Energy in the X-ray radiation pulse versus the length of the second output undulator. Here the output undulator
is 14 cells-long. Grey lines refer to single shot realizations, the black line refers to the average over a hundred realizations.} \label{us2ene}
\end{figure}

\begin{figure}[tb]
\includegraphics[width=1.0\textwidth]{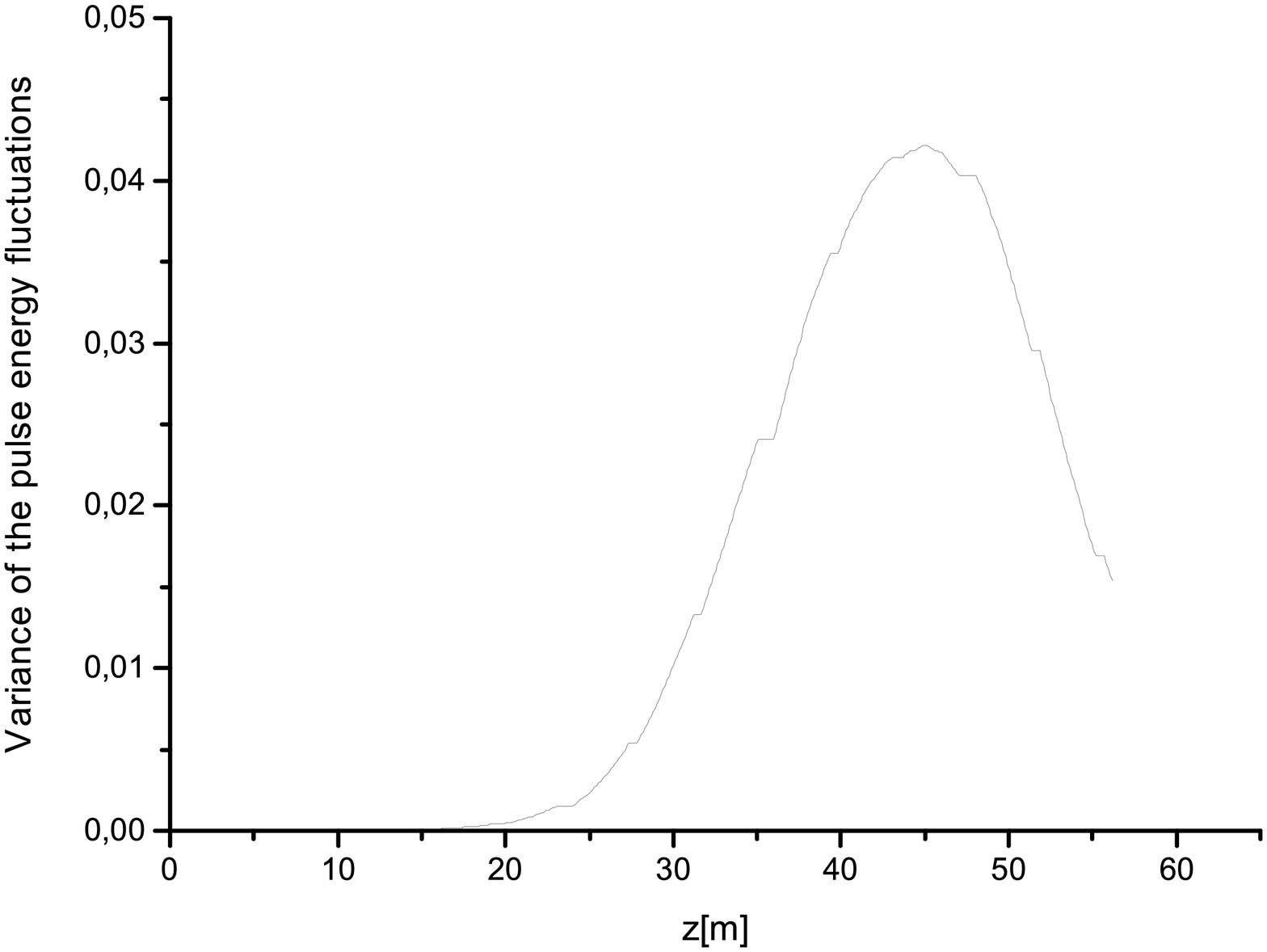}
\caption{Beam doubler setup. Variance of the energy deviation from the average as a function of the
distance inside the second output undulator. Here the output undulator is 14 cells-long.} \label{us2rms}
\end{figure}

\begin{figure}[tb]
\includegraphics[width=1.0\textwidth]{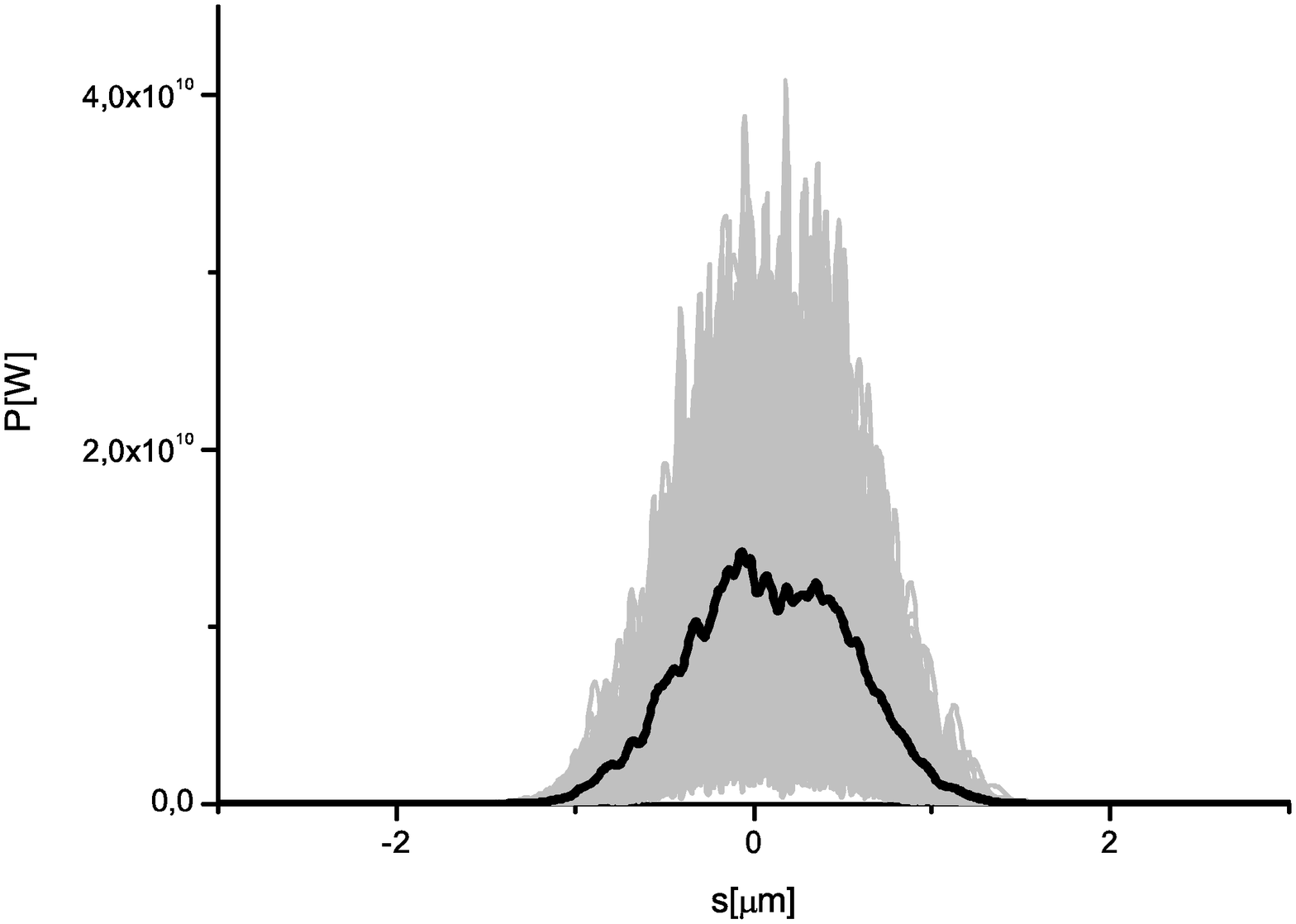}
\caption{Beam doubler setup. Power distribution after the second SASE undulator (14 cells). Grey lines refer to single shot realizations, the black line refers to the average over a hundred realizations. } \label{us2pow}
\end{figure}

\begin{figure}[tb]
\includegraphics[width=1.0\textwidth]{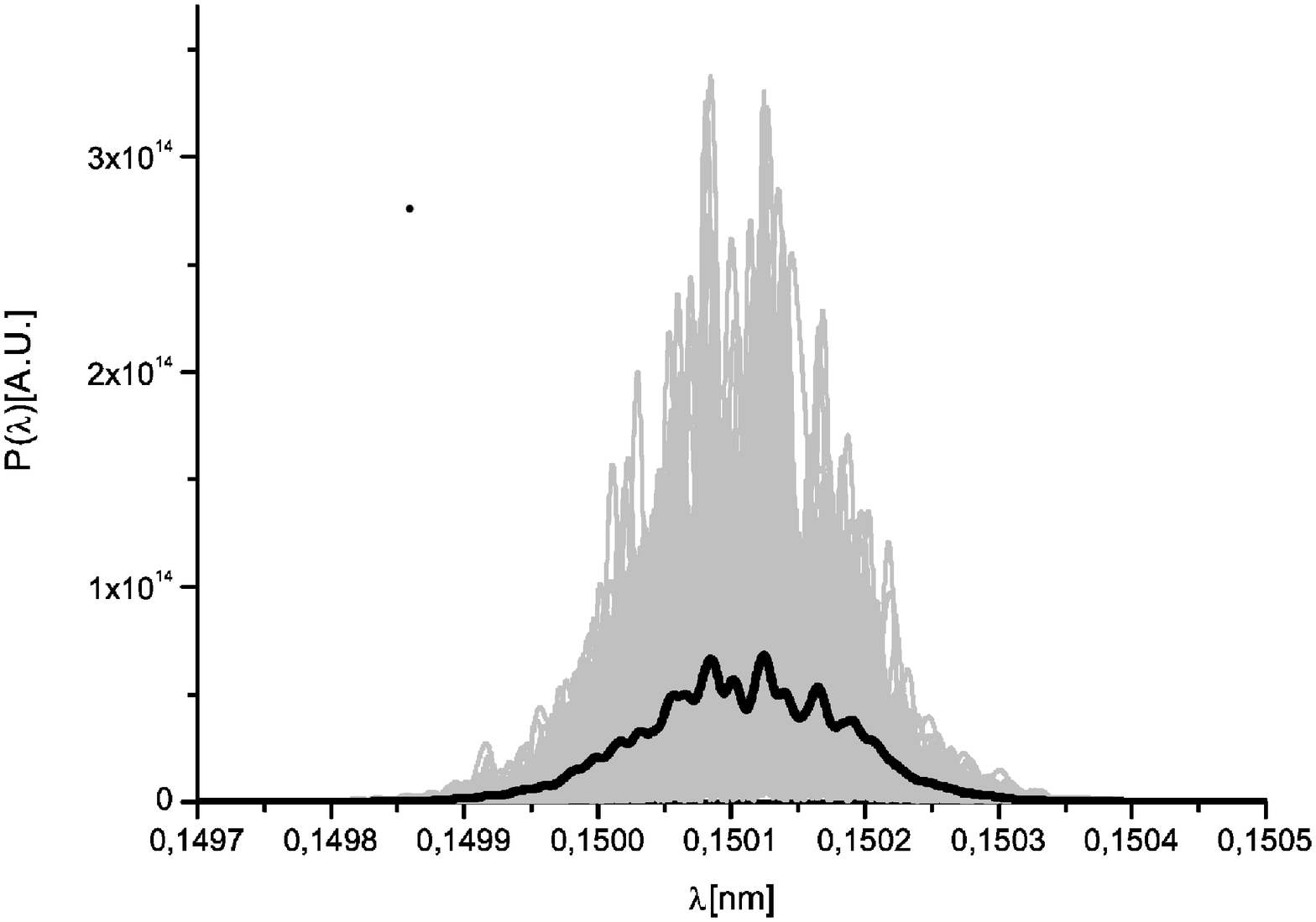}
\caption{Beam doubler setup. Spectrum of the X-ray pulse after the second SASE undulator (14 cells). Grey lines refer to single shot realizations, the black line refers to the average over a hundred realizations.} \label{us2spec}
\end{figure}

The output characteristics, still calculated at $0.15$ nm, are shown in Fig. \ref{us2rms}, Fig. \ref{us2ene}, Fig. \ref{us2pow} and Fig. \ref{us2spec}.

As for the first part of the undulator, Fig. \ref{us2spec} shows a typical SASE spectrum, far from the Fourier limit. Powers are in the $10$ GW-level, Fig. \ref{us2pow}, with energies per pulse in the order of a few tens of microjoules. The poor longitudinal coherence of the SASE pulse results in small energy fluctuations, as can be seen from Fig. \ref{us2rms}.  Fig. \ref{us2rms} also shows that the end of the linear regime is reached since, at it can be seen by inspection, the energy fluctuations drop.


\begin{figure}[tb]
\includegraphics[width=1.0\textwidth]{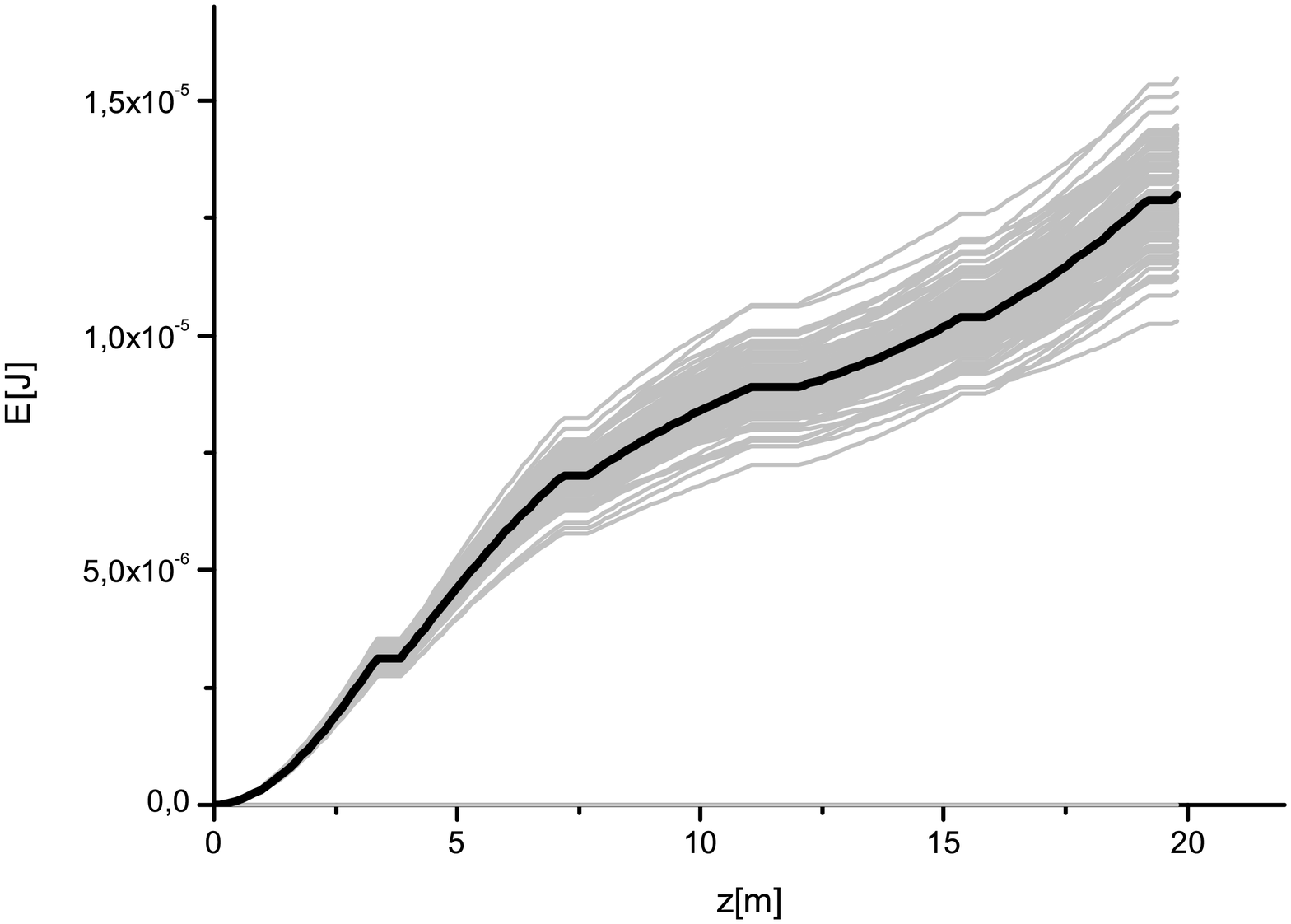}
\caption{Beam doubler setup. Energy in the X-ray radiation pulse versus the length of the SHAB section. Here the SHAB section 5 cells-long. Grey lines refer to single shot realizations, the black line refers to the average over a hundred realizations.} \label{shabene}
\end{figure}

\begin{figure}[tb]
\includegraphics[width=1.0\textwidth]{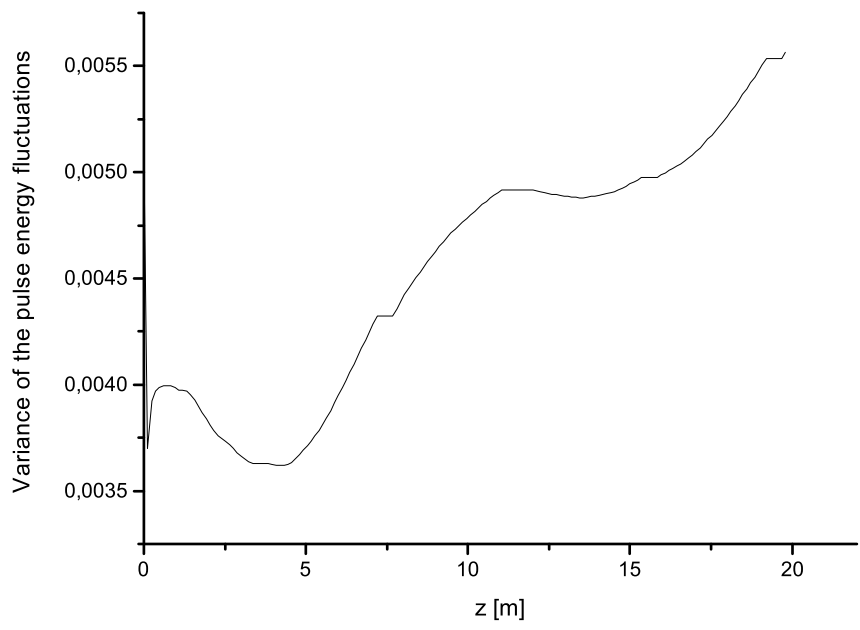}
\caption{Beam doubler setup. Variance of the energy deviation from the average as a function of the distance inside the SHAB section. Here the SHAB section is 5 cells-long.} \label{shabrms}
\end{figure}

\begin{figure}[tb]
\includegraphics[width=1.0\textwidth]{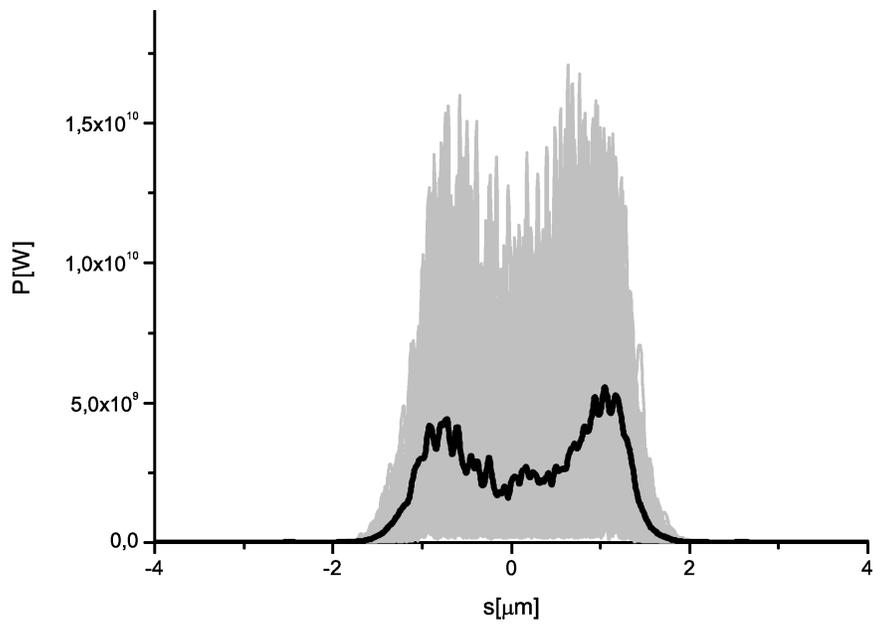}
\caption{Beam doubler setup. Second Harmonic power distribution after the SHAB section (5 cells). Grey lines refer to single shot realizations, the black line refers to the average over a hundred realizations.} \label{shabpow}
\end{figure}

\begin{figure}[tb]
\includegraphics[width=1.0\textwidth]{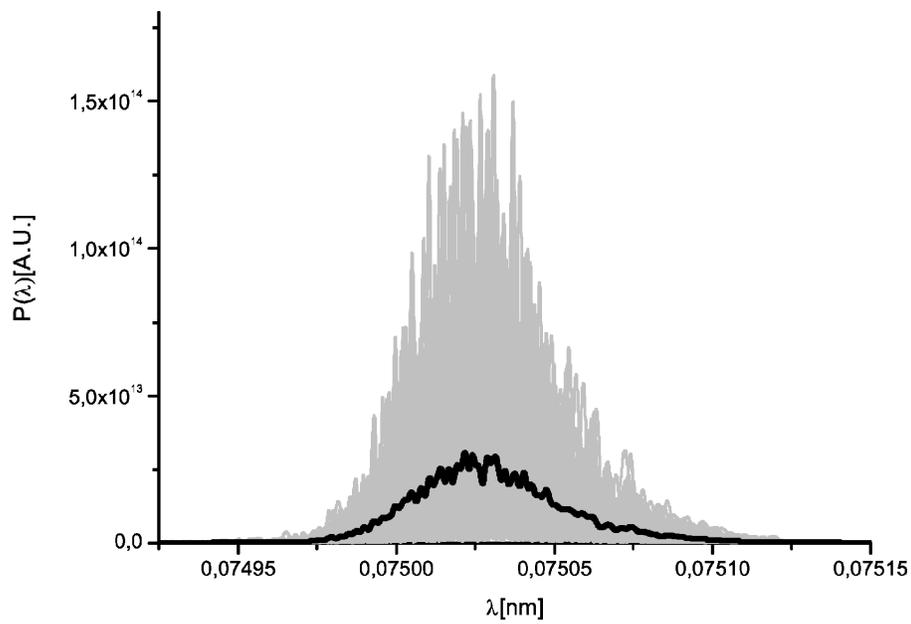}
\caption{Beam doubler setup. Spectrum of the X-ray pulse after the SHAB section (5 cells). Grey lines refer to single shot realizations, the black line refers to the average over a hundred realizations.} \label{shabspec}
\end{figure}

The last five segments in the second undulator part are used as a SHAB, and tuned to the second harmonic of the fundamental. The SHAB takes advantage of the density modulation at the second harmonic developed along the first undulator part. The output characteristics, calculated at $0.075$ nm, are shown in Fig. \ref{shabrms}, Fig. \ref{shabene}, Fig. \ref{shabpow} and Fig. \ref{shabspec}.

As before, Fig. \ref{shabspec} shows a typical SASE spectrum, far from the Fourier limit. Powers are in the GW-level, Fig. \ref{shabpow}, with energies per pulse in the order of ten microjoules. The poor longitudinal coherence of the SASE spectrum results in small energy fluctuations, as can be seen from Fig. \ref{shabrms}.

\clearpage
\section{\label{SHAB} Feasibility study for a self-seeding setup with SHAB}

As explained before, an alternative scheme to that considered in Section \ref{feas} consists in exploiting the self-seeding scheme. This amounts to deactivating the first two segments in the first undulator part, in order to let the first undulator part to lase in the linear regime, so that the self-seeding scheme can be implemented. The mirror setup is now substituted by a wake monochromator, see \cite{OURL}.

\subsection{Undulator Tapering}

\begin{figure}[tb]
\includegraphics[width=1.0\textwidth]{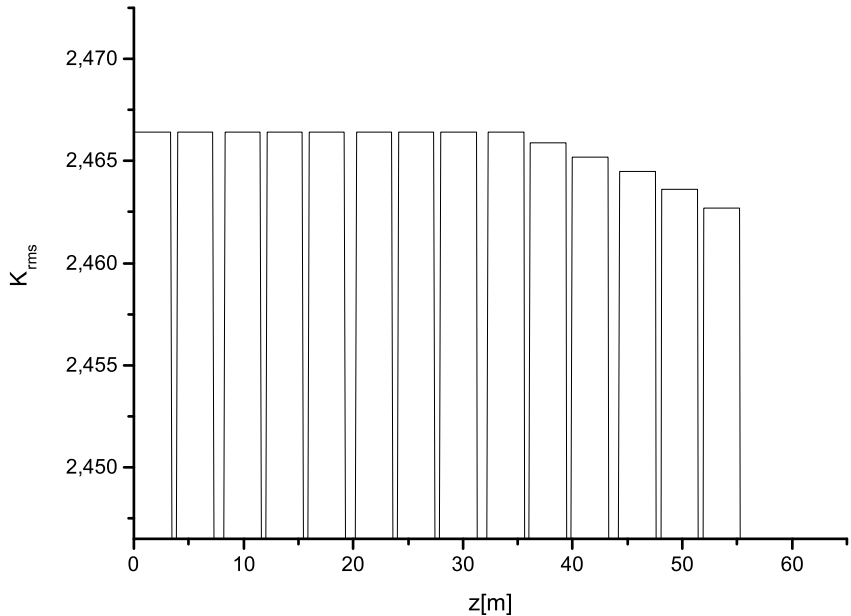}
\caption{Self-seeding setup. Taper configuration for the high power mode of
operation at 0.15 nm.} \label{taplaw}
\end{figure}

\begin{figure}[tb]
\includegraphics[width=1.0\textwidth]{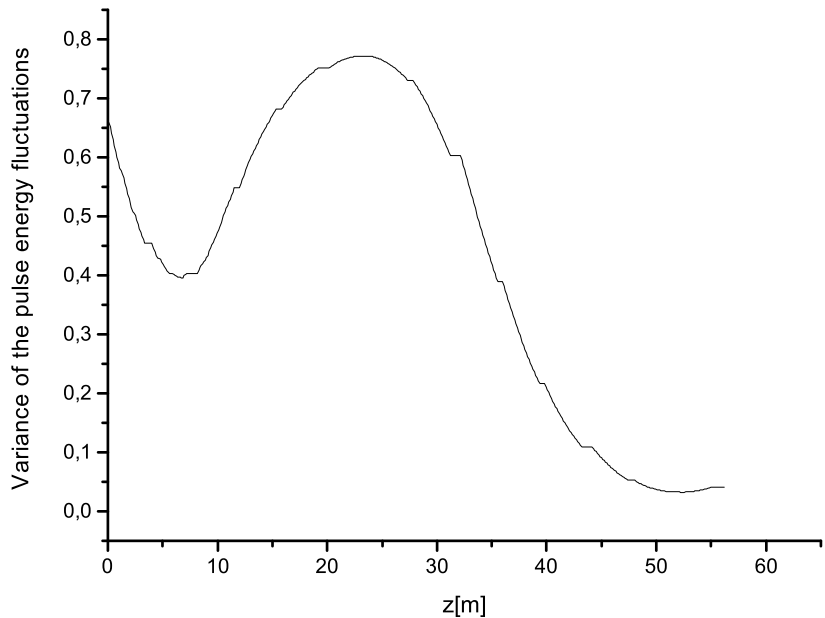}
\caption{Self-seeding setup. First harmonic. Variance of the energy deviation from the average as a function of the distance inside the tapered output undulator. Here the output undulator is constituted by 9 uniform segments and 5 tapered segments, Fig. \ref{taplaw}.} \label{taprms}
\end{figure}

\begin{figure}[tb]
\includegraphics[width=1.0\textwidth]{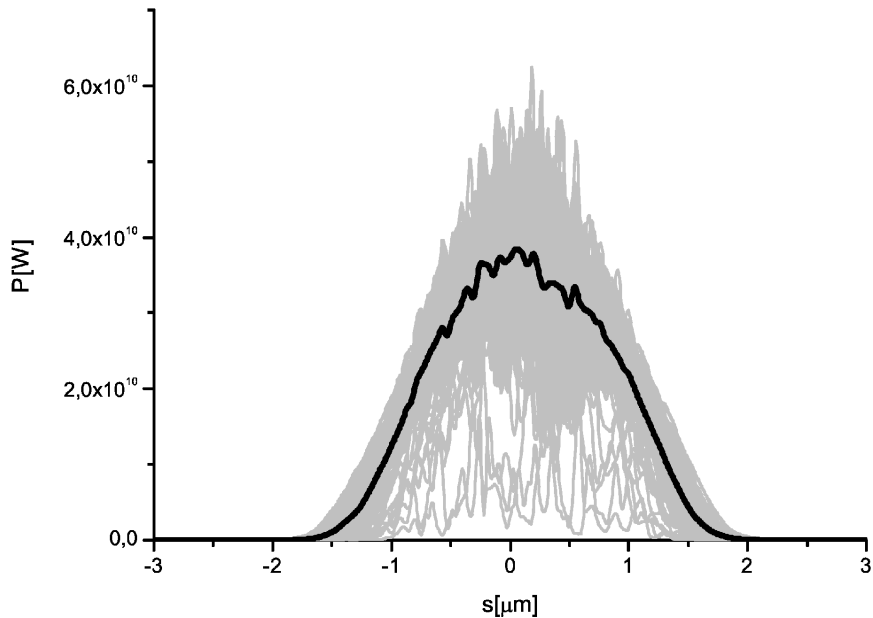}
\caption{Self-seeding setup. First harmonic. Power distribution after the tapered output undulator. Grey lines refer to single shot realizations, the black line refers to the average over a hundred realizations.} \label{tappow}
\end{figure}

\begin{figure}[tb]
\includegraphics[width=1.0\textwidth]{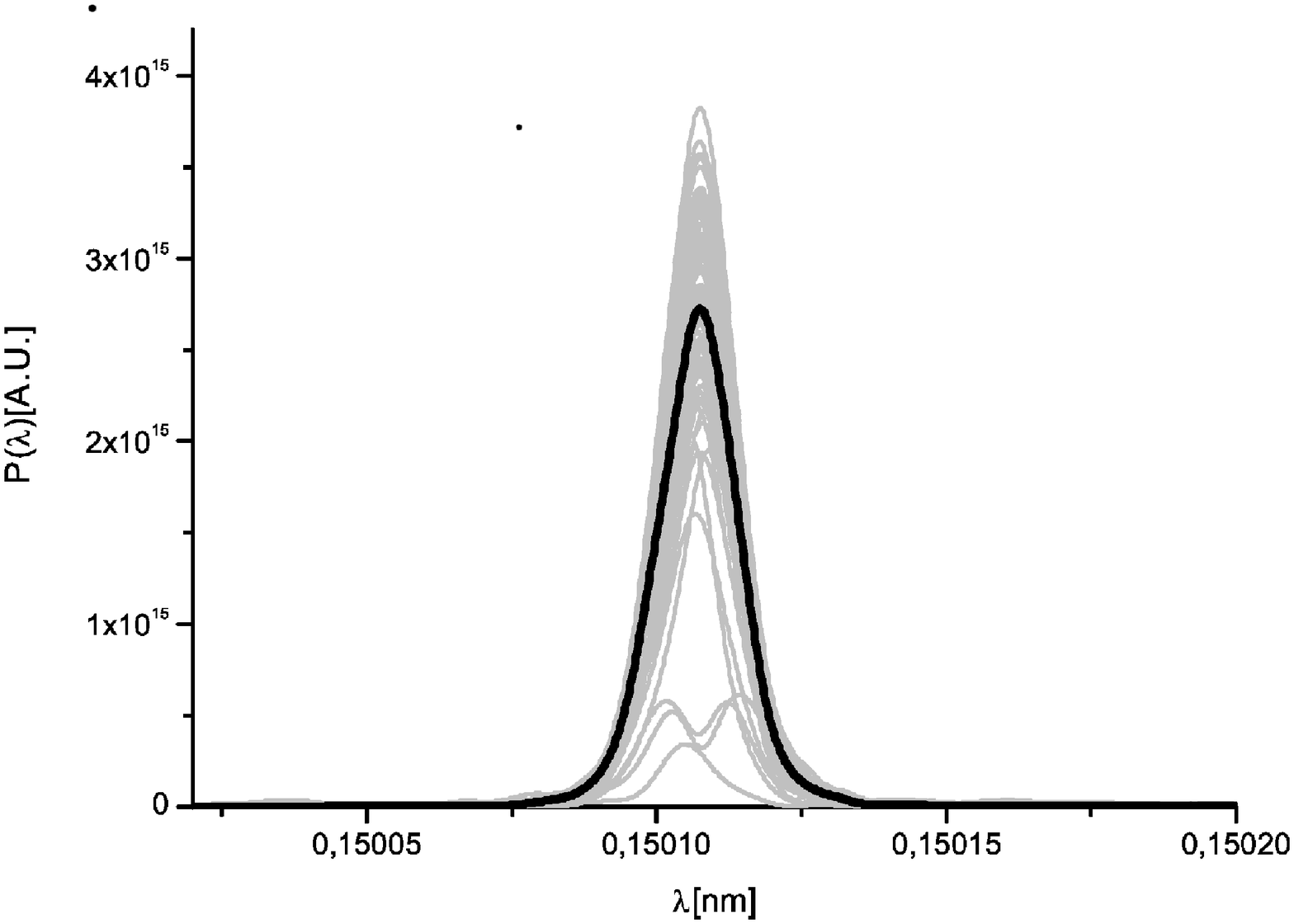}
\caption{Self-seeding setup. Spectrum of the X-ray pulse after the tapered output undulator. Grey lines refer to single shot realizations, the black line refers to the average over a hundred realizations.} \label{tapspec}
\end{figure}

The second (output undulator) can be partly tapered in order to increase the output power. The output power is optimized when the output undulator is tapered, segment by segment,  according to the law in Fig. \ref{taplaw}. The output characteristics, calculated at $0.15$ nm, are shown in Fig. \ref{taprms}, Fig. \ref{tapene}, Fig. \ref{tappow} and Fig. \ref{tapspec}.

Fig. \ref{tapspec} shows the effect of using the wake monochromator, which yields an nearly Fourier limited spectrum. Powers are in the tens of GW-level, Fig. \ref{tappow}, with energies per pulse in the order of hundreds microjoules. The self-seeding scheme results in the typical fluctuation pattern shown in Fig. \ref{taprms}. As discussed in previous papers (see, e.g. \cite{OURL}), at the beginning of the undulator the fluctuations of energy per pulse apparently drop. This can be explained considering the fact that the Genesis output consists of the total power integrated over the full grid up to an artificial boundary, i.e. there is no spectral selection. Therefore, our calculations include a relatively large spontaneous emission background, which has a much larger spectral width with respect to the amplification bandwidth and which fluctuates negligibly from shot to shot. Since there is a long lethargy of the seeded radiation at the beginning of the FEL amplifier, one observes an apparent decrease of fluctuations. Then, when lethargy ends, the seed pulse gets amplified and fluctuations effectively return to about the same level as after monochromator.

\begin{figure}[tb]
\includegraphics[width=1.0\textwidth]{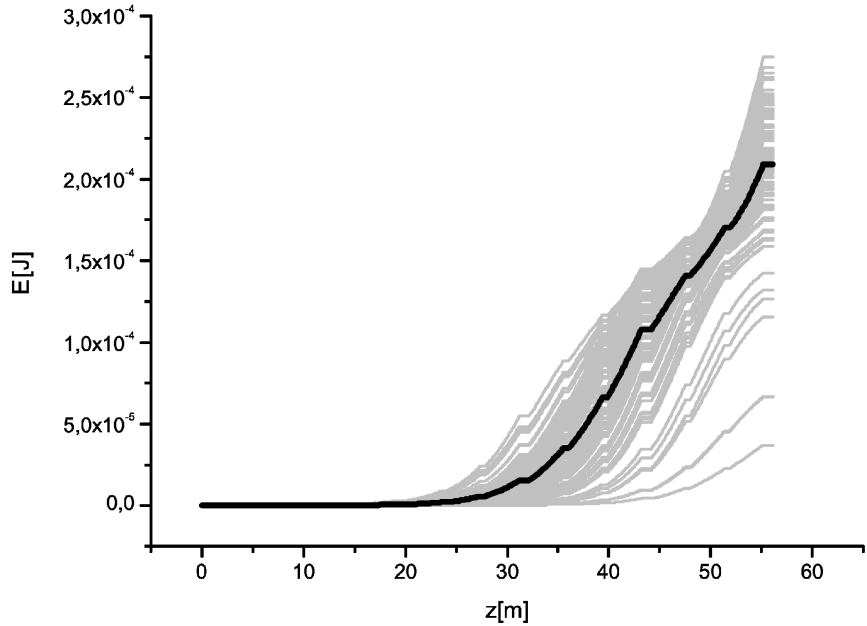}
\caption{Self-seeding setup. First harmonic. Energy in the X-ray radiation pulse versus the length of the tapered output undulator. Here the output undulator is constituted by 9 uniform segments and 5 tapered segments, Fig. \ref{taplaw}. Grey lines refer to single shot realizations, the black line refers to the average over a hundred realizations.} \label{tapene}
\end{figure}


\subsection{Second Harmonic output}

\begin{figure}[tb]
\includegraphics[width=1.0\textwidth]{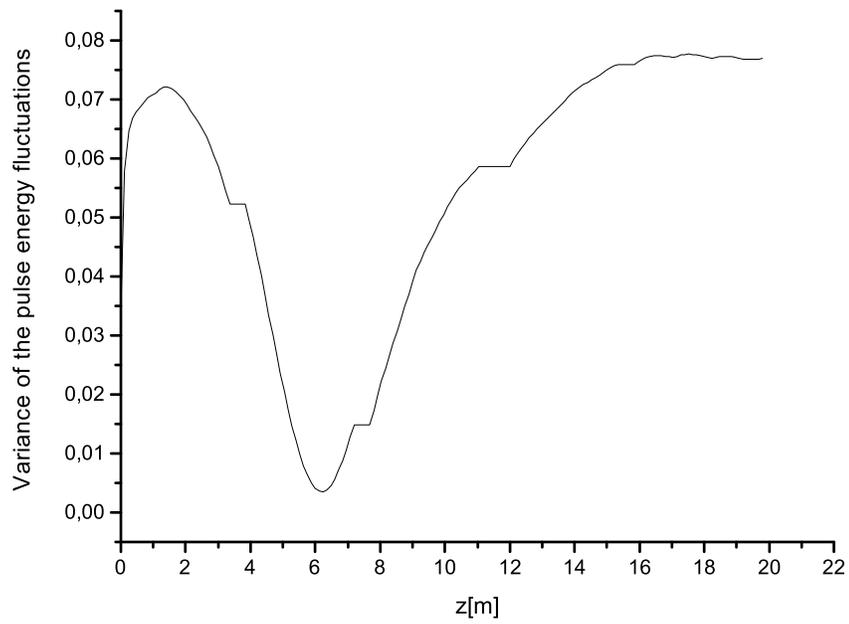}
\caption{Self-seeding setup. Variance of the energy deviation from the average as a function of the
distance inside the second harmonic output undulator. Here the output undulator is constituted by 5 segments.} \label{2habrms}
\end{figure}

\begin{figure}[tb]
\includegraphics[width=1.0\textwidth]{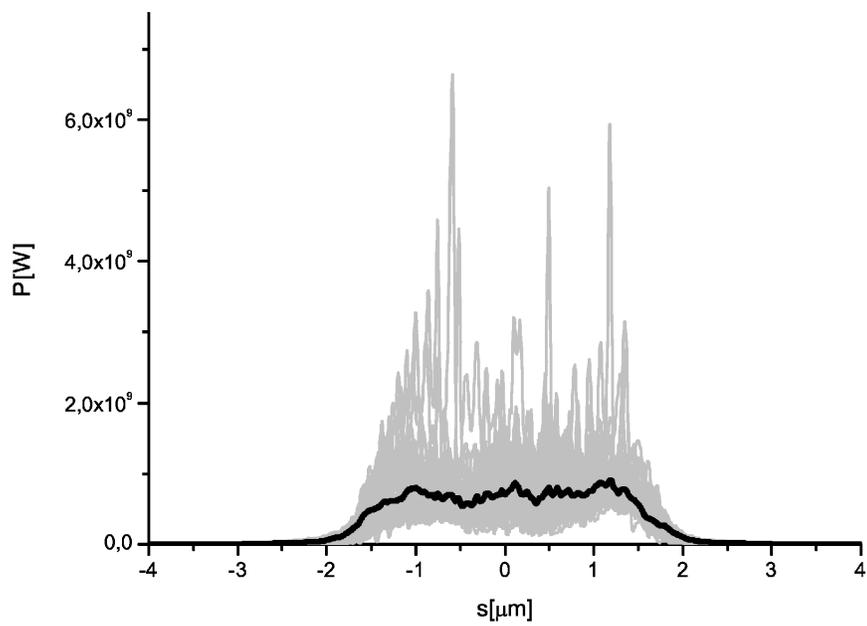}
\caption{Self-seeding setup. Second Harmonic power distribution after the last five cells. Grey lines refer to single shot realizations, the black line refers to the average over a hundred realizations.} \label{2habpow}
\end{figure}

\begin{figure}[tb]
\includegraphics[width=1.0\textwidth]{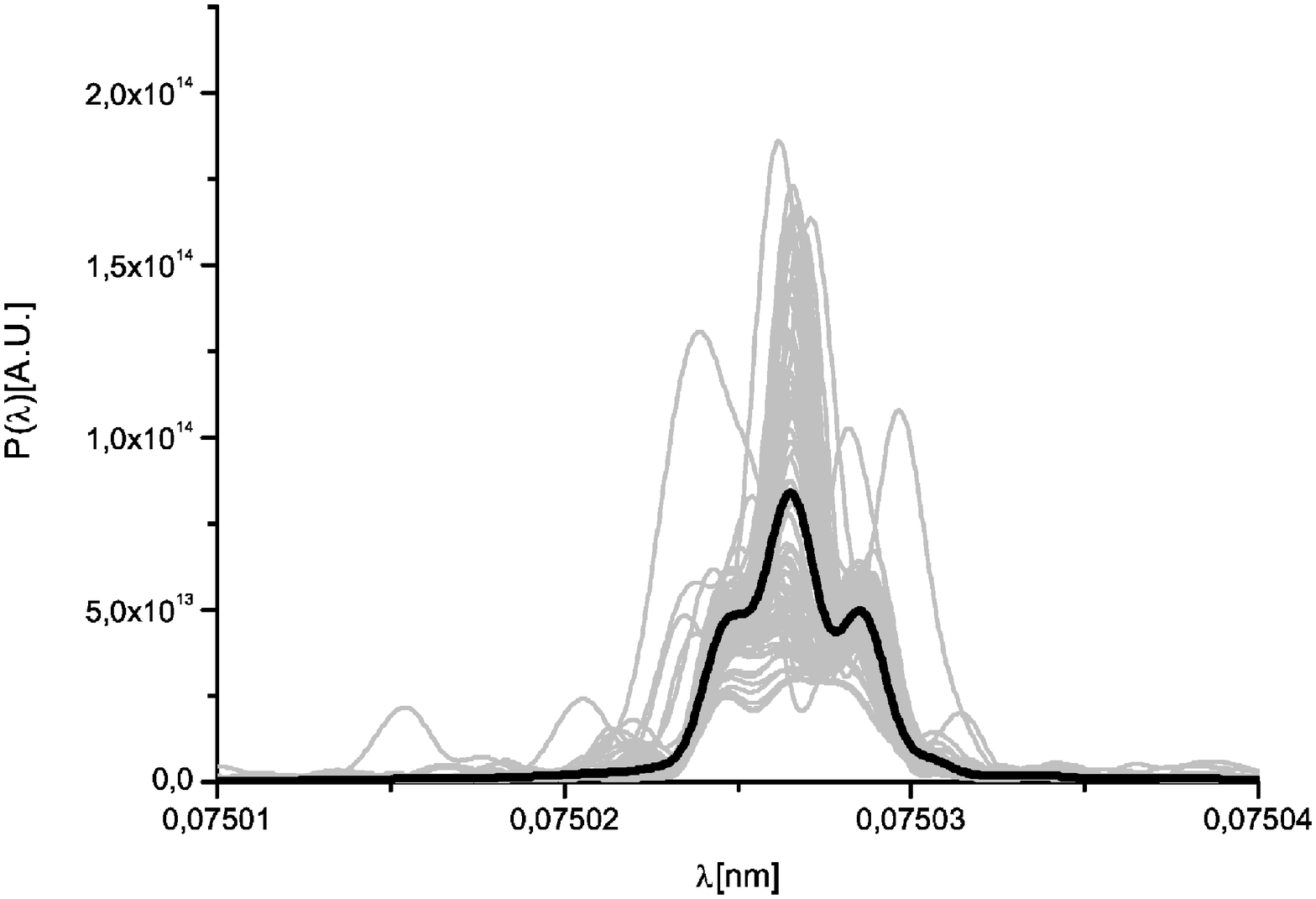}
\caption{Self-seeding setup. Spectrum of the X-ray pulse after the last five cells, tuned at the second harmonic. Grey lines refer to single shot realizations, the black line refers to the average over a hundred realizations.} \label{2habspec}
\end{figure}

Finally, the output from the second-harmonic undulator is shown in Fig. \ref{2habrms}, Fig. \ref{2habpow} and Fig. \ref{2habspec}. The spectrum still shows a good level of monochromatization, Fig. \ref{2habspec}, and the power level is still in the GW level, Fig. \ref{2habpow}. The fluctuation pattern typical of self-seeding schemes is shown in Fig. \ref{2habrms}.

\begin{figure}[tb]
\includegraphics[width=1.0\textwidth]{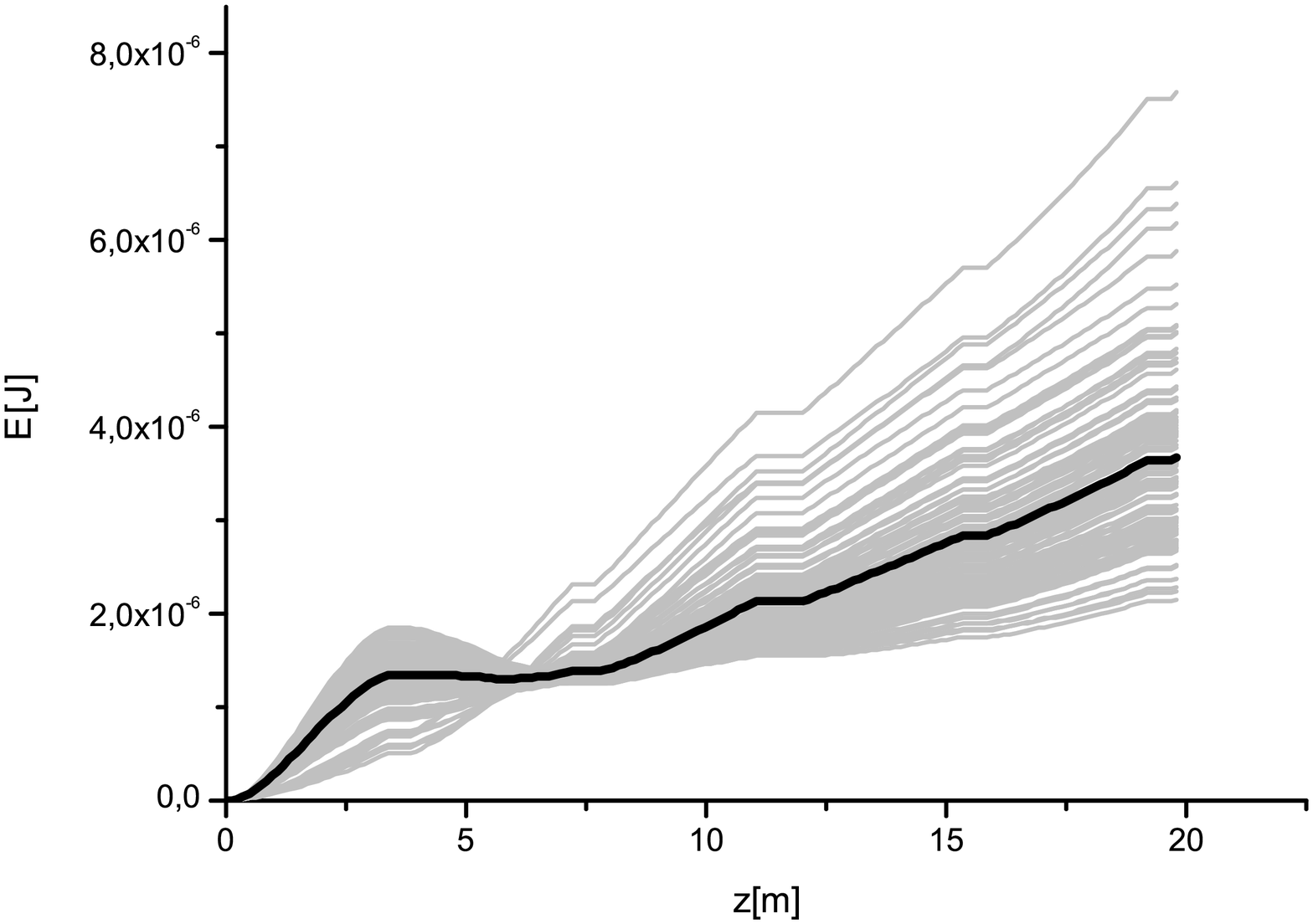}
\caption{Self-seeding setup. Energy in the X-ray radiation pulse versus the length of the second harmonic output undulator. Here the output undulator is constituted by  5  segments. Grey lines refer to single shot realizations, the black line refers to the average over a hundred realizations.} \label{tapene}
\end{figure}

\section{\label{conc} Conclusions}

In this paper we continue to exploit the flexibility of the magnetic chicane setup at LCLS baseline proposed in \cite{OURL} discussing additional options for multiple user operations. The possibility to increase user access is follows from the length of the LCLS baseline undulator, which is about two times longer than what is needed to reach saturation at $0.15$ nm wavelength. Quite surprisingly, the user capacity can be increased with the help of an almost trivial setup, composed by two components: a short magnetic chicane, and an insertable X-ray mirrors pair for X-ray beam separation. It can be straightforwardly installed in the baseline undulator and is safe, in the sense that it guarantees the baseline mode operation. We propose to install the chicane at a special position, after $13$ cells.  In this case we keep all previous options and add, at the same time, the possibility of serving two users with SASE radiation originating from two different parts of the undulator.

This paper also describes an efficient way to extend the LCLS
capabilities to $6$ users working in parallel in the hard X-ray wavelength range, both in the near and far experimental halls. The
high monochromatization of the output radiation constitutes the key for reaching such result and is achieved with self-seeding. In order to exploit the self-seeding option we simply activate $11$ cells out of $13$. Thus, our scheme is based, as for beam doubler, on the introduction of a  short magnetic chicane, which is extremely flexible and can be straightforwardly adopted for self-seeding with  single crystal monochromator.

For the first time we simulated self-seeding with SHAB, and studied the properties of the second harmonic in this case. We demonstrated that this method can produce intense monochromatic radiation both at the first and at the second harmonic. We optimized the first harmonic power and demonstrated that even with a small number  of cells it is still possible to reach 40 GW power. For the second harmonic we obtained a GW level with a narrow bandwidth 0.007 $\%$. We propose to use the self-seeding mode of operation with SHAB for enabling multi-user mode of operation. For this we propose to use rapidly
and frequently switching crystal deflectors. This is possible due to the relatively large photon energy (16 keV), resulting in a small absorption.   Switching of deflectors will be provided by a small tilting of the crystal. Such rotation can be realized similarly as for phase retarders for polarization switching in synchrotron radiation facilities. Routine parameters of existing synchrotron radiation facilities (e.g. PETRA III) allow for a 100 Hz repetition rate for the rotation cycle, and a rotation error  of $0.1 \mu$rad.

As a result, in this paper we discuss about two separate options for multi-user operation. The first deals with the possibility of two users in whole wavelength range. The second allows for $6$ users, but for the hard X-ray range only. The proposed photon beam distribution system would allow to switch the X-ray beam quickly between  6 experiment in order to make an efficient use of the LCLS source.

\section{Acknowledgements}

We are grateful to Massimo Altarelli, Reinhard Brinkmann, Serguei
Molodtsov and Edgar Weckert for their support and their interest
during the compilation of this work.

\end{document}